\documentclass[twocolumn,showpacs,prb,aps,psfrag]{revtex4} 
\usepackage{amsmath,amssymb,epsf,bm}
\usepackage{dcolumn}
\usepackage{graphicx} 
\usepackage[usenames,dvipsnames]{color}
\usepackage{psfrag,latexsym} 
\usepackage{epstopdf}
\usepackage[caption=false]{subfig}
\usepackage{comment}
\usepackage{ulem}

\begin{document}

\title{Spin Waves in Ferromagnetic Insulators Coupled via a Normal Metal}

\author{Hans Skarsv{\aa}g}
\email{hans.skarsvag@ntnu.no}
\affiliation{Department of Physics, Norwegian University of Science and
Technology, NO-7491 Trondheim, Norway}

\author{Andr\'{e} Kapelrud}
\affiliation{Department of Physics, Norwegian University of Science and
Technology, NO-7491 Trondheim, Norway}

\author{Arne Brataas}
\affiliation{Department of Physics, Norwegian University of Science and
Technology, NO-7491 Trondheim, Norway}

\date{\today}

\begin{abstract}
Herein, we study the spin-wave dispersion and dissipation in a ferromagnetic insulator--normal metal--ferromagnetic insulator system. Long-range dynamic coupling because of spin pumping and spin transfer lead to collective magnetic excitations in the two thin-film ferromagnets. In addition, the dynamic dipolar field contributes to the interlayer coupling. By solving the Landau-Lifshitz-Gilbert-Slonczewski equation for macrospin excitations and the exchange-dipole volume as well as surface spin waves, we compute the effect of the dynamic coupling on the resonance frequencies and linewidths of the various modes. The long-wavelength modes may couple acoustically or optically. In the absence of spin-memory loss in the normal metal, the spin-pumping-induced Gilbert damping enhancement of the acoustic mode vanishes, whereas the optical mode acquires a significant Gilbert damping enhancement, comparable to that of a system attached to a perfect spin sink. The dynamic coupling is reduced for short-wavelength spin waves, and there is no synchronization. For intermediate wavelengths, the coupling can be increased by the dipolar field such that the modes in the two ferromagnetic insulators can couple despite possible small frequency asymmetries. The surface waves induced by an easy-axis surface anisotropy exhibit much greater Gilbert damping enhancement. These modes also may acoustically or optically couple, but they are unaffected by thickness asymmetries. 
\end{abstract}


\maketitle
\section{Introduction}

The dynamic magnetic properties of thin-film ferromagnets have been extensively studied for several decades.\cite{Kittel:prev48,Damon:chemsol61} Thin-film ferromagnets exhibit a rich variety of spin-wave modes because of the intricate interplay among the exchange and dipole interactions and the material anisotropies. In ferromagnetic insulators (FIs), these modes are especially visible; the absence of disturbing electric currents leads to a clear separation of the magnetic behavior. Furthermore, the dissipation rates in insulators are orders of magnitude lower than those in their metallic counterparts; these low dissipation rates enable superior control of travelling spin waves and facilitate the design of magnonic devices. \cite{Cherepanov:PR93} 

In spintronics, there has long been considerable interest in giant magnetoresistance, spin-transfer torques, and spin pumping in hybrid systems of normal metals and metallic ferromagnets (MFs).\cite{Binasch:prb89,Berger:prb96,Slonczewski:jmmm96,Brataas:nmat12} The experimental demonstration that spin transfer and spin pumping are also active in normal metals in contact with insulating ferromagnets has generated a renewed interest in and refocused attention on insulating ferromagnets, of which yttrium iron garnet (YIG) continues to be the prime example.\cite{Kajiwara:nat10,Sandweg:apl10,Sandweg:prl11,Heinrich:prl11,Vilela-Leao:apl11,Rezende:apl13,Burrowes:apl12,Pirro:apl14,Nakayama:prl13,Hahn:prb13,Vlietstra:prb13,Brataas:physics13} In ferromagnetic insulators, current-induced spin-transfer torques from a neighboring {\it{normal metal}} (NM) that exhibits out-of-equilibrium spin accumulation may manipulate the magnetization of the insulator  and excite spin waves.\cite{Kajiwara:nat10,klein:arx14,Xiao:prl12} The out-of-equilibrium spin accumulation of the normal metal may be induced via the spin Hall effect or by currents passing through other adjacent conducting ferromagnets. Conversely, excited spin waves pump spins into adjacent NMs, and this spin current may be measured  in terms of the inverse spin Hall voltages or by other conducting ferromagnets.\cite{Kajiwara:nat10,Sandweg:apl10,Sandweg:prl11,Heinrich:prl11,Vilela-Leao:apl11,Rezende:apl13,Burrowes:apl12} The magnetic state may also be measured via the spin Hall magnetoresistance.\cite{Lu:prb13,Nakayama:prl13,Hahn:prb13,Vlietstra:prb13,Chen:prb13,Brataas:physics13} Because of these developments, magnetic information in ferromagnetic insulators  may be electrically injected, manipulated, and detected. Importantly, an FI-based spintronic device may efficiently transport electric information carried by spin waves over long distances\cite{Pirro:apl14} without any excessive heating. The spin-wave decay length can be as long as centimeters in YIG films.\cite{Schneider:apl08} These properties make FI--NM systems ideal devices for the exploration of novel spintronic phenomena and possibly also important for future spintronic applications. Magnonic devices also offer advantages such as rapid spin-wave propagation, frequencies ranging from GHz to THz, and the feasibility of creating spin-wave logic devices and magnonic crystals with tailored spin-wave dispersions.\cite{Kruglyak:jphysd2010}

To utilize the desirable properties of FI--NM systems, such as the exceptionally low magnetization-damping rate of FIs, it is necessary to understand how the magnetization dynamics couple to spin transport in adjacent normal metals. The effective damping of the uniform magnetic mode of a thin-film FI is known to significantly increase when the FI is placed in contact with an NM. This damping enhancement is caused by the loss of angular momentum  through spin pumping.\cite{Urban:prl01,Mizukami:jmmm02,Tserkovnyak:prl02,Brataas:prb02,Tserkovnyak:rmp05} Recent theoretical work has also predicted the manner in which the Gilbert damping for other spin-wave modes should become renormalized. \cite{Kapelrud:prl13}  For long-wavelength spin waves, the Gilbert damping enhancement is twice as large for transverse volume waves as for the macrospin mode, and for surface modes, the enhancement can be ten times stronger or more. Spin pumping has been demonstrated, both experimentally\cite{Sandweg:apl10} and theoretically,\cite{Kapelrud:prl13} to be suppressed for short-wavelength exchange spin waves. 

A natural next step is to investigate the magnetization dynamics of more complicated FI--NM heterostructures. In ferromagnetic metals, it is known that spin pumping and spin-transfer torques generate a long-range dynamic interaction between magnetic films separated by normal metal layers.\cite{Heinrich:prl03} The effect of this long-range dynamic interaction on homogeneous macrospin excitations can be measured by ferromagnetic resonance. The combined effects of spin pumping and spin-transfer torque lead to an appreciable increase in the resonant linewidth when the resonance fields of the two films are far apart and to a dramatic narrowing of the linewidth when the resonant fields approach each other.\cite{Heinrich:prl03} This behavior occurs because the excitations in the two films couple acoustically (in phase) or optically (out of phase). We will demonstrate that similar, though richer because of the complex magnetic modes, phenomena exist in magnetic insulators. 

In the present paper, we investigate the magnetization dynamics in a thin-film stack consisting of two FIs that are in contact via an NM. The macrospin dynamics in a similar system with metallic ferromagnets have been studied both theoretically and experimentally.\cite{Heinrich:prl03} We expand on that work by focusing on inhomogeneous magnetization excitations in FIs.  

For long-wavelength spin waves travelling in-plane in a ferromagnetic thin film, the frequency as a function of the in-plane wave number $Q$ strongly depends on the direction of the external magnetic field with respect to the propagation direction. If the external field is in-plane and the spin waves are travelling parallel to this direction, the waves have a negative group velocity. Because the magnetization precession amplitudes are usually evenly distributed across the film in this geometry, these modes are known as backward volume magnetostatic spin waves (BVMSW). Similarly, spin waves that correspond to out-of-plane external fields are known as forward volume magnetostatic spin waves (FVMSW), i.e., the group velocity is positive, and the precession amplitudes are evenly distributed across the film. When the external field is in-plane and perpendicular to the propagation direction, the precession amplitudes of the spin waves become inhomogeneous across the film, experiencing localization to one of the interfaces. These spin waves are thus known as magnetostatic surface spin waves (MSSW).\cite{Kalinikos:jphys,Serga:jphysd}

When two ferromagnetic films are coupled via a normal metal, the spin waves in the two films become coupled through two different mechanisms. First, the dynamic, nonlocal dipole-dipole interaction causes an interlayer coupling to arise that is independent of the properties of the normal metal. This coupling is weaker for larger thicknesses of the normal metal. Second, spin pumping from one ferromagnetic insulator induces a spin accumulation in the normal metal, which in turn gives rise to a spin-transfer torque on the other ferromagnetic insulator, and vice versa. This dynamic coupling, is in contrast to the static exchange coupling \cite{grunberg:prl89} rather long-ranged and is limited only by the spin-diffusion length. This type of coupling is known to strongly couple the macrospin modes.  When two ferromagnetic films become coupled, the characterization of the spin waves in terms of FVMSW, BVMSW, and MSSW still holds, but the dispersion relations are modified. It is also clear that the damping renormalization caused by spin pumping into the NM may differ greatly from that in a simpler FI$|$N bilayer system. To understand this phenomenon, we perform a detailed analytical and numerical analysis of a trilayer system, with the hope that our findings may be used as a guide for experimentalists.

This paper is organized as follows. Section~\ref{sec:equmot} introduces the model. The details of the dynamic dipolar field are discussed, and the boundary conditions associated with spin pumping and spin transfer at the FI$|$N interfaces are calculated. Sec.~\ref{analytical} provides the analytical solutions of these equations in the long-wavelength regime dominated by the dynamic coupling attributable to spin pumping and spin transfer. To create a more complete picture of the dynamic behavior of this system, we perform a numerical analysis for the entire spin-wave spectrum of this system, which is presented in Sec.~\ref{numerical}. We conclude our work in Sec.~\ref{conclusion}. 

\section{Equations of motion}\label{sec:equmot}
\begin{figure}[!ht]
    \subfloat[\label{fig:coord2D}]{
 \includegraphics[width=0.15\textwidth,trim = 10mm 20mm 10mm 20mm]{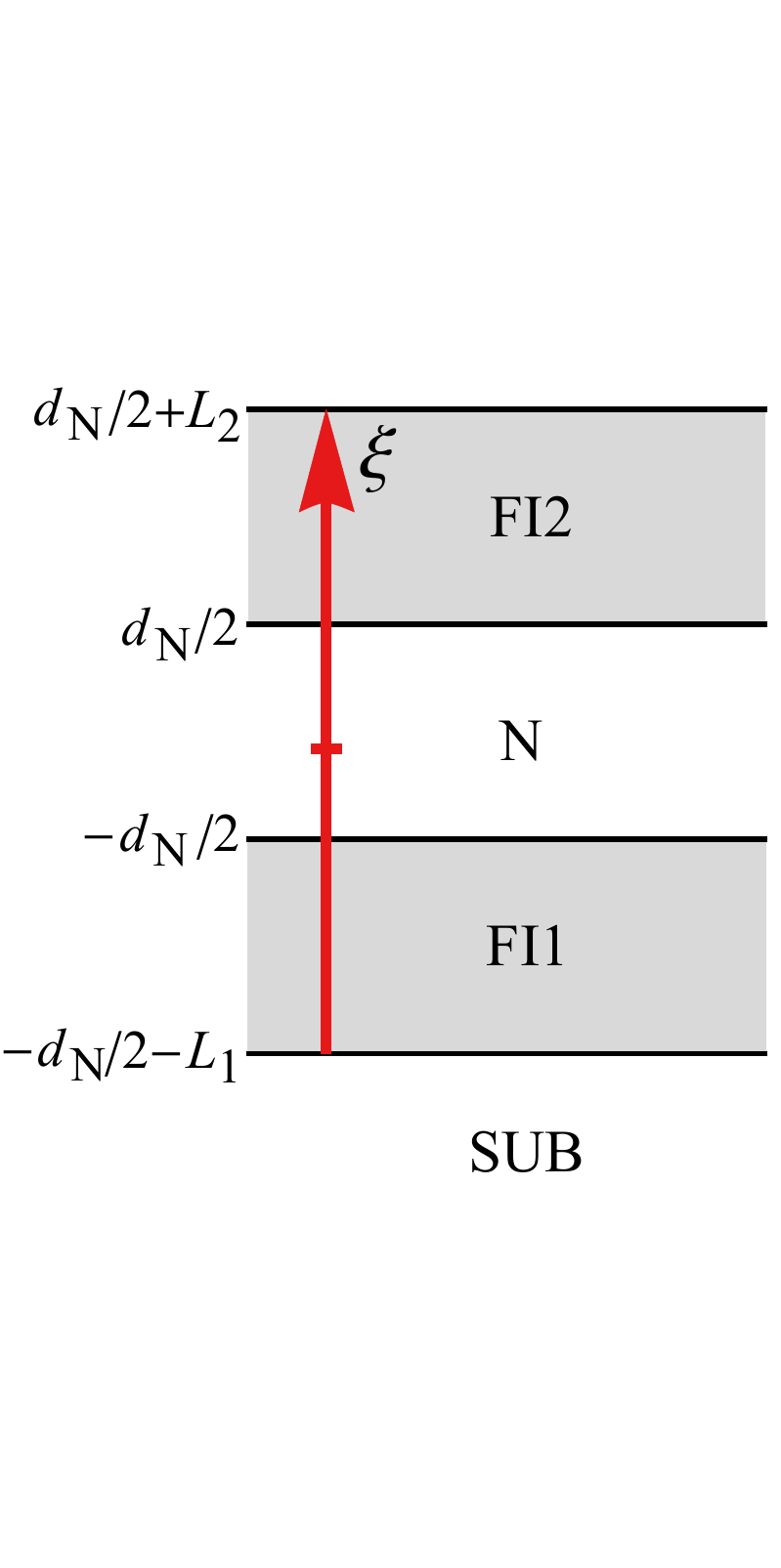}
    }
    \hfill
    \subfloat[\label{fig:coord3D}]{
      \includegraphics[width=0.2\textwidth,trim =25mm -10mm 5mm 40mm]{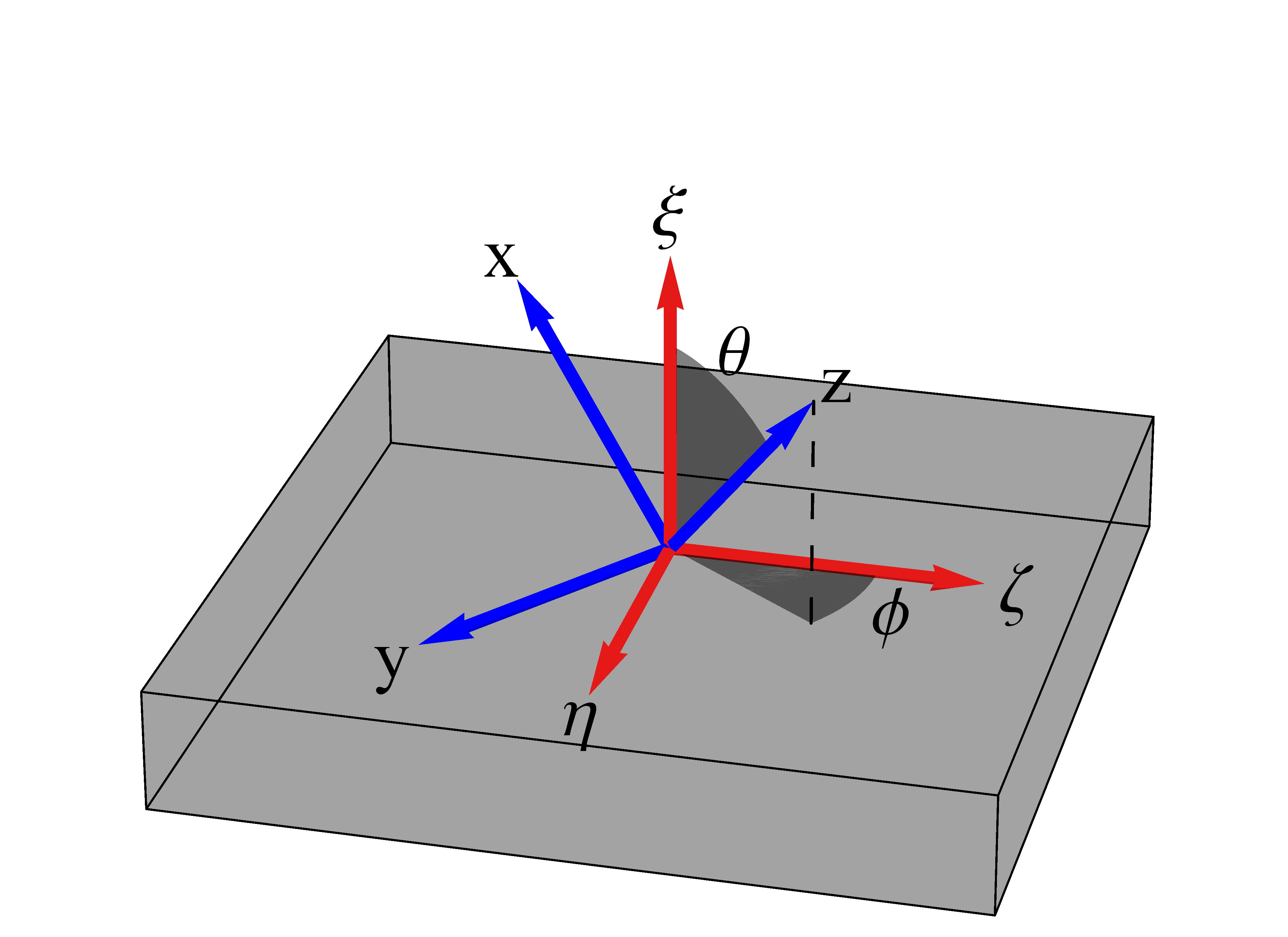}
    }
	\caption{(Color online) a) A cross section of the FI1$|$N$|$FI2 heterostructure.
The ferromagnetic insulators FI1 and FI2 are in contact via the normal metal N.
The transverse coordinate $\xi$ is indicated along with the thicknesses $L_1$,
$d_\text{N}$, and $L_2$ of FI1, N, and FI2, respectively. b) The coordinate
system of the internal field (blue) with respect to the coordinate system of the
FI1$|$N$|$FI2 structure (red). $\theta$ denotes the angle between the film
normal and the internal field, and $\phi$ is the angle between the in-plane component of the magnetic 
field and the in-plane wave vector.}
    \label{fig:coordinates}
  \end{figure}

Consider a thin-film heterostructure composed of two ferromagnetic insulators (FI1 and FI2) that are in electrical contact via an NM layer. The ferromagnetic insulators FI1 and FI2 may have different thicknesses and material properties. We denote the thicknesses by $L_1$, $d_\text{N}$, and $L_2$ for the FI1, NM, and FI2 layers, respectively (see Fig.~\ref{fig:coordinates}(a)). The in-plane coordinates are $\zeta, \eta$, and the transverse coordinate is $\xi$ (see Fig.~\ref{fig:coordinates}(b)). We will first discuss the magnetization dynamics in isolated FIs and will then incorporate the spin-memory losses and the coupling between the FIs via spin currents passing through the NM. 
\subsection{Magnetization Dynamics in Isolated FIs}
The magnetization dynamics in the ferromagnetic insulators can be described by using the Landau-Lifshitz-Gilbert (LLG) equation,
\begin{equation}
	\dot{\mathbf{M}}_i=-\gamma  \mathbf{M}_i\times\mathbf{H}_\text{eff}+\alpha \mathbf{M}_i\times \dot{\mathbf{M}}_i,
\end{equation}
where $\mathbf{M}_i$ is the unit vector in the direction of the magnetization in layer $i=1,2$, $\gamma$ is the gyromagnetic ratio, $\alpha$ is the dimensionless damping parameter, and $\mathbf{H}_\text{eff}$ is the space-time-dependent  effective magnetic field. The effective magnetic field is 
\begin{equation}
	\mathbf{H}_\text{eff}=\mathbf{H}_\text{int}+\mathbf{h}_\text{ex}+\mathbf{h}_\text{d}+\mathbf{h}_\text{surface},
\end{equation}
where $\mathbf{H}_\text{int}$ is the internal field attributable to an external magnetic field
and the static demagnetization field, $\mathbf{h}_\text{ex}=2A\nabla^2\mathbf{M}/M_S$ is
the exchange field ($A$ is the exchange constant), $\mathbf{h}_\text{d}$ is the
dynamic demagnetization field, and 
\begin{equation}\label{surface anisotropy}
	\mathbf{h}_{\text{surface}}=
	\frac{2K_{S}}{M_{S}^2}(\mathbf{M}_i\cdot \hat{\mathbf{n}})\delta(\xi-\xi_i)\hat{\mathbf{n}}
\end{equation}
is the surface anisotropy field located at the FI$|$N interfaces. In this work, $\mathbf{h}_\text{surface}$ is assumed to exist only at the FI$|$N interfaces and not at the interfaces between the FIs and the substrate or vacuum. It is straightforward to generalize the discussion to include these surface anisotropies as well. We consider two scenarios: one with an easy-axis surface anisotropy ($K_S>0$) and one with no surface anisotropy ($K_S=0$). Note that a negative value of $K_S\sim -0.03~\mathrm{erg/cm^2}$, which implies an easy-plane surface anisotropy, has also been observed for sputtered YIG$|$Au bilayers.\cite{Heinrich:private} In general, the effective field $\mathbf{{H}_\text{eff}}$ may differ in the two FIs. We assume the two FIs consist of the same material and consider external fields that are either in-plane or out-of-plane. Furthermore, we consider devices in which the internal magnetic fields in the two FI layers are aligned and of equal magnitude. 

In equilibrium, the magnetization inside the FIs is oriented along the internal magnetic field, $\mathbf{M}_i=\mathbf{M}_0$. In the linear response regime, $\mathbf{M}_i=\mathbf{M}_0+\mathbf{m}_i$, where the first-order correction $\mathbf{m}_i$ is small and perpendicular to $\mathbf{M}_0$. The magnetization vanishes outside of the FIs. Because the system is translationally invariant in the $\eta$ and $\zeta$ directions, we may, without loss of generality, assume that $\mathbf{m}$ consists of plane waves travelling in the $\zeta$ direction,
\begin{equation}
	\mathbf{m}_i(\zeta,\eta,\xi)=\mathbf{m}_{iQ}(\xi)e^{i(\omega t-Q\zeta)} \,.
\end{equation}
Linearizing Maxwell's equations in $\mathbf{m}_i$ implies that the dynamic dipolar field must be of the same form,
\begin{equation}
 \mathbf{h}_\text{d}(\zeta,\eta,\xi)=\mathbf{h}_{\text{d}Q}(\xi)e^{i(\omega
  t-Q\zeta)}\,.
\end{equation}
Furthermore, the total dipolar field (the sum of the static and the dynamic dipolar fields) must satisfy Maxwell's equations, which, in the magnetostatic limit, are
\begin{subequations}\label{eq:maxwell}
\begin{eqnarray}
		\nabla \cdot \left(\mathbf{h}_\text{d}+4\pi M_S \mathbf{m}\right)&=&0,\\
		\nabla \times \mathbf{h}_\text{d}&=&0,
\end{eqnarray}
\end{subequations} 
with the boundary equations
\begin{subequations}\label{eq:maxwellboundary}
\begin{eqnarray}
	(\mathbf{h}_\text{d}+4\pi M_S\mathbf{m})_{\perp,\text{in}}&=&(\mathbf{h}_\text{d})_{\perp,\text{out}},\\
	(\mathbf{h}_\text{d})_{\parallel,\text{in}}&=&(\mathbf{h}_\text{d})_{\parallel,\text{out}},\,
\end{eqnarray}
\end{subequations} 
where the subscript in (out) denotes the value on the FI (NM, vacuum or substrate) side of the FI interface and $\perp$ ($\parallel$) denotes the component(s) perpendicular (parallel) to the FI--NM interfaces. Solving Maxwell's equations  \eqref{eq:maxwell} with the boundary conditions of Eq.~(\ref{eq:maxwellboundary}) yields\cite{Kalinikos:jphys}
\begin{equation}
	\mathbf{h}_{\text{d}Q}(\xi)=\int
	\mathrm{d}\xi'\hat{G}\left(\xi-\xi'\right)\mathbf{m}_{Q}(\xi'), \label{eq:dyndip}
\end{equation}
where $\hat{G}(\mathbf{r}-\mathbf{r}')$ is a 3$\times$3 matrix acting on $\mathbf{m}$ in the $(\eta, \zeta, \xi)$ basis,
\begin{equation}\label{kalinikos}
	\hat{G}(\xi)=\left(\begin{matrix}
		G^\text{P}(\xi)-\delta(\xi) & 0 &-iG^\text{Q}(\xi)\\
		0&0&0\\
		-iG^\text{Q}(\xi)&0& -G^\text{P} (\xi)
	\end{matrix}\right).
\end{equation}
Here, $G^\text{P}(\xi)=Qe^{-Q\mid \xi \mid}/2$, and $G^\text{Q}(\xi)=-\text{sign}(\xi)G^{\text{P}}$. Note that the dynamic dipolar field of Eq.~\eqref{eq:dyndip} accounts for both the interlayer and intralayer dipole-dipole couplings  because the magnetization varies across the two magnetic insulator bilayers and vanishes outside these materials.
\begin{figure}[!ht]
	\includegraphics[width=0.40\textwidth,trim = 15mm 65mm 15mm 65mm]{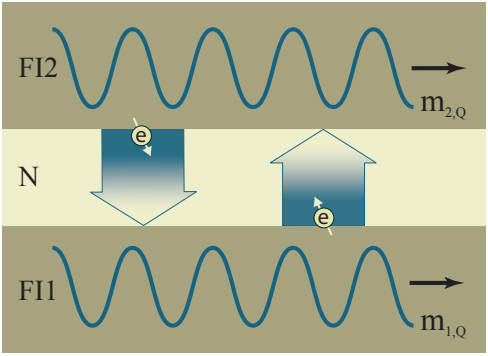}
	\caption{(Color online) Two coupled spin waves with amplitude $\mathbf{m}_{1Q}$ in ferromagnet FI1 and amplitude $\mathbf{m}_{2Q}$ in ferromagnet FI2. The spin-waves inject a spin current into the normal metal (NM) via spin pumping. In the NM, the spins diffuse and partially relax, inducing a spin accumulation therein. In turn, the spin accumulation causes spin-transfer torques to arise on FI1 and FI2. The combined effect of spin transfer and spin pumping leads to a dynamic exchange coupling that, together with the dynamic demagnetization field, couples the spin waves in the two FIs.}
	\label{fig:SWinteraction}
\end{figure}

It is now convenient to perform a transformation from the $\zeta$-$\eta$-$\xi$ coordinate system defined by the sample geometry to the $x$-$y$-$z$ coordinate system defined by the internal field (see Fig.~\ref{fig:coordinates}(b)). In the linear response regime, the dynamic magnetization $\mathbf{m}_i$ lies in the $x$-$y$ plane, and the linearized equations of motion become\cite{Kalinikos:jphys}
\begin{widetext}
\begin{equation}\label{matrix LLG}
\left[i\omega\left(\begin{matrix} \alpha & -1 \\ 1 & \alpha
\end{matrix}\right)+\openone\left(\omega_H+\frac{2A}{M_S}\left[Q^2-\frac{\mathrm{d}^2}{\mathrm{d}\xi^2}\right]\right)\right] \mathbf{m}_{iQxy}(\xi)=\sum_{i=1}^2\int \mathrm{d}\xi' \hat{\mathcal{G}}_{xy} (\xi-\xi')\mathbf{m}_{iQxy}(\xi').
\end{equation}
\end{widetext}
Here, $\mathbf{m}_{iQxy}=(m_{iQx},m_{iQy})$ is the Fourier transform of the dynamic component of the magnetization in the $x$-$y$ plane and $\hat{\mathcal{G}}_{xy}(\xi)$ is the 2$\times$2 matrix that results from rotating $\hat{G}(\xi)$ into the $x$-$y$-$z$ coordinate system (see Appendix \ref{app: dipole tensor}), and considering only the $xx$, $xy$, $yx$ and $yy$-components.

\subsection{Boundary Conditions and Spin Accumulation}
The linearized equations of motion (\ref{matrix LLG}) must be supplemented with boundary conditions for the dynamic magnetization at the FI$|$N interfaces. A precessing magnetization at the FI$|$N boundaries injects a spin-polarized current, $\mathbf{j}^{\text{SP}}$, into the NM, an effect known as \textit{spin pumping}. \cite{Tserkovnyak:prl02,Brataas:prb02,Tserkovnyak:rmp05,Kajiwara:nat10} The emitted spin currents at the lower and upper interfaces ($i=1,2$) are
\begin{equation}\label{spinpumpingcurrent}
\mathbf{j}^\text{SP}_i=\left.\frac{\hbar}{e}g_\perp \mathbf{M}_i\times \dot{\mathbf{M}}_i\right|_{\xi=\xi_i},
\end{equation}
where $\xi_i=\mp d_\text{N}/2$ at the lower and upper interfaces, respectively, and $g_\perp$ is the real part of the transverse spin-mixing conductance per unit area.\cite{Brataas:prl00} We disregard the imaginary part of the spin-mixing conductance because it has been found to be small at FI$|$N interfaces.\cite{Althammer:prb13} The reciprocal effect of spin pumping is spin transfer into the FIs because of a spin accumulation $\boldsymbol{\mu}_S$ in the NM. In the normal metal at the lower and upper interfaces ($i$=1,2), the associated spin-accumulation-induced spin current is 
\begin{equation}\label{spintransfercurrent}
\mathbf{j}^\text{ST}_ i=\left. -\frac{1}{e}g_\perp \mathbf{M}_i\times \left(\mathbf{M}_i\times \boldsymbol{\mu}_S\right) \right|_{\xi=\xi_i}.
\end{equation}
The signs of the pumped and spin-accumulation-induced spin currents in Eqs.~\eqref{spinpumpingcurrent} and \eqref{spintransfercurrent} were chosen such that they are positive when there is a flow of spins from the NM toward the FIs. 

The pumped and spin-accumulation-induced spin currents of Eqs.~\eqref{spinpumpingcurrent} and (\ref{spintransfercurrent}) lead to magnetic torques acting on the FI interfaces. The torques that correspond to the spin pumping and spin transfer localized at the FI$|$N interfaces are
\begin{subequations}
\begin{eqnarray}
\boldsymbol{\tau}_i^\text{SP}&=&\frac{\gamma \hbar^2}{2e^2}g_\perp \delta(\xi-\xi_i)\mathbf{M}_i\times \dot{\mathbf{M}}_i,\\
\boldsymbol{\tau}^\text{ST}_i&=&-\frac{\gamma \hbar}{2e^2}g_\perp\mathbf{M}_i\times(\mathbf{M}_i\times \boldsymbol{\mu}_{S})\delta(\xi-\xi_i),
\end{eqnarray}
\end{subequations}
respectively. In the presence of spin currents to and from the normal metal, the magnetization dynamics in the FIs is then governed by the modified Landau-Lifshitz-Gilbert-Slonczewski (LLGS) equation,
\begin{equation}\label{LLGS}
\dot{\mathbf{M}}=-\gamma  \mathbf{M}_i\times\mathbf{H}_\text{eff}+\alpha \mathbf{M}_i\times \dot{\mathbf{M}}_i+\sum_{i=1,2}\boldsymbol{\tau}^\text{SP}_i+\boldsymbol{\tau}^\text{ST}_i.
\end{equation}
By integrating Eq.~\eqref{LLGS} over the FI$|$N interfaces and the interfaces between the FI and vacuum/substrate, we find that $\mathbf{m}_i$ must satisfy the boundary conditions \cite{Xiao:prl12, Kapelrud:prl13}
\begin{subequations}\label{boundary condition}
\begin{eqnarray}\label{boundary condition a}
\left(\pm L_i\frac{\mathrm{d} \mathbf{m}_i}{\mathrm{d} \xi}+\chi_i\left[\dot{\mathbf{m}}_i-\frac{1}{\hbar}\mathbf{M}_0\times\boldsymbol{\mu}\right]\right.\hspace{20mm}\vspace{-4mm}&\nonumber\\ \vspace{-4mm}
\left. \left.+\frac{L_i K_S}{A} \cos\left(2\theta\right)\mathbf{m}_i\right)_x\right|_{\xi=\mp d_\text{N}/2}=0,&
\end{eqnarray}    
\begin{eqnarray}\label{boundary condition b}
\left(\pm L_i\frac{\mathrm{d} \mathbf{m}_i}{\mathrm{d} \xi}+\chi_i\left[\dot{\mathbf{m}}_i-\frac{1}{\hbar}\mathbf{M}_0\times\boldsymbol{\mu}\right]\right.\hspace{20mm}&\nonumber\\
\left.\left.+\frac{L_i K_s}{A} \cos^2\left(\theta\right)\mathbf{m}_i\right)_y\right|_{\xi=\mp d_\text{N}/2}=0,&
\end{eqnarray}   
\begin{eqnarray}\label{boundary condition c}
\left.\frac{\mathrm{d} \mathbf{m}_1}{\mathrm{d} \xi}\right|_{\xi=-d_\text{N}/2-L_1}=0, \hspace{4mm} \left.\frac{\mathrm{d} \mathbf{m}_2}{\mathrm{d} \xi}\right|_{\xi=d_\text{N}/2+L_2}=0.\hspace{-3mm}
\end{eqnarray}
\end{subequations}
Here, we have introduced the timescale $\chi_i=L_i\hbar^2 g_\perp/4Ae^2$. The subscripts $x$ and $y$ in Eqs.~(\ref{boundary condition a}) and (\ref{boundary condition b}) denote the $x$ and $y$ components, respectively. In our expressions for the boundary conditions (\ref{boundary condition}), we have also accounted for the possibility of a surface anisotropy arising from the effective field described by Eq.~\eqref{surface anisotropy}, where $K_S>0$ indicates an easy-axis surface anisotropy (EASA). The boundary conditions of Eq.~\eqref{boundary condition}, in combination with the transport equations in the NM , which we will discuss next, determine the spin accumulation in the NM and the subsequent torques caused by spin transfer.

In the normal metal, the spins diffuse,  creating a spatially dependent spin-accumulation potential $\boldsymbol{\mu}_Q$, and they relax on the spin-diffusion length scale $l_\text{sf}$.  The spin accumulation for an FI$|$N$|$FI system has been calculated in the macrospin model.\cite{Skarsvaag:14} The result of this calculation can be directly generalized to the present situation of spatially inhomogeneous spin waves by replacing the macrospin magnetization in each layer with the interface magnetization and substituting the spin-diffusion length with a wave-vector-dependent effective spin-diffusion length $l_\text{sf}\rightarrow \tilde{l}_\text{sf}(Q)$ such that
\begin{equation}
\label{lsfeff}
\begin{split}
\boldsymbol{\mu}_Q=-\frac{\hbar}{2} \mathbf{M}_0\times \left[ (\dot{\mathbf{m}}_{Q}(\xi_1)+\dot{\mathbf{m}}_{Q}(\xi_2))\Gamma_1\left(\xi\right)\right.\\
\left.-(\dot{\mathbf{m}}_{Q}(\xi_1)-\dot{\mathbf{m}}_{Q}(\xi_2))\Gamma_2\left(\xi\right)\right].
\end{split}
\end{equation}
See Appendix \ref{app: spin acc} for the details of the functions $\Gamma_1$ and $\Gamma_2$. The effective spin-diffusion length is found by Fourier transforming the spin-diffusion equation (see Appendix \ref{app: lsf}), resulting in
\begin{equation}\label{eq:lsfeff}
\tilde{l}_\text{sf}=l_\text{sf}/\sqrt{1+(Ql_\text{sf})^2}.
\end{equation}

We thus have all the necessary equations to describe the linear response dynamics of spin waves in the FI1$|$N$|$FI2 system. We now provide analytical solutions of the spin-wave modes in the long-wavelength limit and then complement these solutions with an extensive numerical analysis that is valid for any wavelength. 

\section{Analytic solutions for the spin wave spectrum}
\label{analytical}

The effect that the exchange and dipolar fields have on the spin-wave spectrum depends on the in-plane wave number $Q$. When $QL_i\ll 1$, the dipolar field dominates over the exchange field. In the opposite regime, when $QL_i \gg 1$, the exchange field dominates over the dipolar field. The intermediate regime is the dipole-exchange regime. Another length scale is set by the spin-diffusion length. When $Ql_\text{sf}\gg1$, the effective spin-relaxation length $\tilde{l}_\text{sf}$ of Eq.~(\ref{eq:lsfeff}) becomes small, and the NM acts as a perfect spin sink. In this case, only the relatively short-ranged dipolar field couples the FIs. We therefore focus our attention on the dipole-dominated regime, in which the interchange of spin information between the two FIs remains active.

In the limit $QL_i\ll 1$, the magnetization is homogeneous in the in-plane direction. We may then use the ansatz that the deviation from equilibrium is a sum of transverse travelling waves. Using the boundary conditions on the outer boundaries of the stack, Eq.~\eqref{boundary condition c}, we find 
\begin{equation}\label{eq:amplitudevec}
\mathbf{m}_{iQxy}(\xi)=\left(\begin{matrix} X_i \\ Y_i \end{matrix} \right) \cos \left\{ k_i \left[\xi\pm (L_i+\frac{d_\text{N}}{2}) \right]\right\},
\end{equation}
where $i=1$ when
$\xi$ is inside FI1 and $i=2$ when $\xi$ is inside FI2. $k_1$ and $k_2$ are the out-of-plane wave vectors of the lower and upper films, respectively. The eigenfrequencies of Eq.~\eqref{matrix LLG} depend on $k_i$. To first order in the damping parameter $\alpha$, we have
\begin{widetext}
\begin{equation}\label{bulk dispersion}
\omega(k_i)= \omega_M \left[ \pm\sqrt{\left(\frac{\omega_H}{\omega_M} +\frac{A}{2\pi M_S^2}k^2_i\right) \left(\frac{\omega_H}{\omega_M} +\frac{A}{2\pi M_S^2}k^2_i+\sin ^2\theta \right) }+i\alpha \left(\frac{\omega_H}{\omega_M} +\frac{A}{2\pi M_S^2}k^2_i+\frac{1}{2}\sin ^2\theta \right)\right].
\end{equation}
\end{widetext}
We can, without loss of generality, consider only those frequencies that have a positive real part. The eigenfrequency $\omega$ is a characteristic feature of the entire system, so we must require $\omega(k_1)=\omega(k_2)$, which implies that $k_1=\pm k_2$. We will discuss the cases of symmetric ($L_1=L_2$) and asymmetric ($L_1\neq L_2$) geometries separately. 

\subsection{Symmetric FI films without EASA}
\label{symFInoEASA}
Consider a symmetric system in which the FIs are of identical thickness and material properties. We assume that the effect of the EASA is negligible, which is the case for thin films and/or weak surface anisotropy energies such that $K_S L/A\ll 1$, where $L=L_1=L_2$. The other two boundary conditions,
\eqref{boundary condition a} and \eqref{boundary condition b}, couple the amplitude vectors $\left(\begin{matrix} X_1&Y_1 \end{matrix}\right)^\text{T}$ and $\left(\begin{matrix} X_2&Y_2 \end{matrix}\right)^\text{T}$ of Eq.~\eqref{eq:amplitudevec}. A non-trivial solution implies that the determinant that contains the coefficients of the resulting $4\times 4$ matrix equation vanishes. Solving the secular equation, we find the following constraints on $k$,
\begin{subequations}\label{eq:eigenvaluemodes}
\begin{eqnarray}
i\chi_\text{A}\omega_\text{A}=kL\tan(kL),\\
i\chi_\text{O}\omega_\text{O}=kL\tan(kL),
\end{eqnarray}
\end{subequations}
where 
\begin{subequations}\label{eq:eigenvaluecoefficients}
\begin{eqnarray}
\chi_\text{A}=\chi\left(1-\left[1+\frac{2g_\perp l_\text{sf}}{\sigma}\tanh(d_\text{N}/2l_\text{sf})\right]^{-1}\right),\\
\chi_\text{O}=\chi\left(1-\left[1+\frac{2g_\perp l_\text{sf}}{\sigma}\coth(d_\text{N}/2l_\text{sf})\right]^{-1}\right),
\end{eqnarray}
\end{subequations}
and $\chi=L\hbar^2 g_\perp/4Ae^2$. The two solutions correspond to a symmetric mode (acoustic) and an antisymmetric mode (optical). This result can be understood in terms of the eigenvectors that correspond to the eigenvalues of Eqs.~(\ref{eq:eigenvaluemodes}), which are $\mathbf{m}_1=+\mathbf{m}_2$ and $\mathbf{m}_1=-\mathbf{m}_2$ for the acoustic and optical modes, respectively. Typically, because spin pumping only weakly affects the magnetization dynamics, the timescale $\chi$ that is proportional to the mixing conductance $g_\perp$ is much smaller than the FMR precession period. In this limit, $kL \tan (kL)\ll 1$. This result allows us to expand the secular equations (\ref{eq:eigenvaluemodes}) around $kL=n\pi$, where $n$ is an integral number, which yields
\begin{equation}
i\chi_\nu\omega_{\nu,n}\approx(kL+\pi n)kL,
\end{equation}
where $\nu=\text{A},\text{O}$. This result can be reinserted into the bulk dispersion relation of Eq.~\eqref{bulk dispersion}, from which we can determine the renormalization of the Gilbert damping coefficient attributable to spin pumping, $\Delta \alpha$. We define 
\begin{equation}\label{eq:alphadef}
\Delta \alpha=\alpha \left(\text{Im}[\omega^{(\text{SP})}]-\text{Im}[\omega^{(0)}]\right)/\text{Im}[\omega^{(0)}]
\end{equation}
as a measure of the spin-pumping-enhanced Gilbert damping, where $\omega^{(0)}$ and $\omega^{(\text{SP})}$ are the frequencies of the same system without and with spin pumping, respectively.

Similar to the case of a single-layer ferromagnetic insulator,\cite{Kapelrud:prl13} we find that all higher transverse volume modes exhibit an enhanced magnetization dissipation that is twice that of the macrospin mode. The enhancement of the Gilbert damping for the macrospin mode ($n=0$) is
\begin{equation}\label{eq:macroalpha}
\Delta \alpha _{\nu,\text{macro}}=\frac{\gamma\hbar^2 g_\perp}{2LM_Se^2} \frac{\chi_\nu}{\chi},
\end{equation}
and for the other modes, we obtain
\begin{equation}\label{eq:modealpha}
\Delta \alpha _{\nu,n\neq 0}=2\Delta \alpha _{\nu,\text{macro}}.
\end{equation}

Compared with single-FI systems, the additional feature of systems with two FIs is that the spin-pumping-enhanced Gilbert damping differs significantly between the acoustic and optical modes via the mode-dependent ratio $\chi_\nu/\chi$. This phenomenon has been explored both experimentally and theoretically in Ref. \onlinecite{Heinrich:prl03} for the macrospin modes  $n=0$ when there is no loss of spin transfer between the FIs, $l_\text{sf} \rightarrow \infty$. Our results represented by Eqs.~(\ref{eq:macroalpha}) and (\ref{eq:modealpha}) are generalizations of these results for the case of other transverse volume modes and account for spin-memory loss. Furthermore, in Sec. IV, we present the numerical results for the various spin-wave modes when the in-plane momentum $Q$ is finite. When the NM is a perfect spin sink, there is no transfer of spins between the two FIs, and we recover the result for a single FI$|$N system with vanishing back flow, $\chi_\nu\rightarrow \chi$. \cite{Kapelrud:prl13} Naturally, in this case, the FI$|$N$|$FI system acts as two independent FI$|$N systems with respect to magnetization dissipation. The dynamical interlayer dipole coupling is negligible in the considered limit of this section ($QL\ll 1$).

In the opposite regime, when the NM film is much thinner than the spin-diffusion length and the spin conductivity of the NM is sufficiently large such that $g_\perp d_\text{N}/\sigma\ll 1$, then $\chi_\text{A}\rightarrow 0$ and $\chi_\text{O}\rightarrow \chi$.  This result implies that for the optical mode, the damping is the same as for a single FI in contact with a perfect spin sink, even though the spin-diffusion length is very large. The reason for this phenomenon is that when the optical mode is excited, the magnetizations of the two films oscillate out of phase such that one layer acts as a perfect spin sink for the other layer. By contrast, there is no enhancement of the Gilbert damping coefficient for the acoustic mode; when the film is very thin and the magnetizations of the two layers are in phase, there is no net spin flow or loss in the NM film and no spin-transfer-induced losses in the ferromagnets. Finally, when the NM is a poor conductor despite exhibiting low spin-memory loss such that $g_\perp d_\text{N}/\sigma \gg (l_\text{sf}/d_\text{N}) \gg 1$, then $\chi_\nu\rightarrow 0$ because there is no exchange of spin information. For the macrospin modes in the absence of spin-memory loss, these results are in exact agreement with Ref.~\onlinecite{Heinrich:prl03}. Beyond these results, we find that regardless of how much spin memory is lost, it is also the case that in trilayer systems, all higher transverse modes experience a doubling of the spin-pumping-induced damping. Furthermore, these modes can still be classified as optical and acoustic modes with different damping coefficients.

\subsection{Symmetric Films with EASA}
Magnetic surface anisotropy is important when the spin-orbit interaction at the interfaces is strong. In this case, the excited mode with the lowest energy becomes inhomogeneous in the transverse direction. For a finite $K_S$, the equations for the $x$ and $y$ components of the magnetization in the boundary condition \eqref{boundary condition} differ, resulting in different transverse wave vectors for the two components, $k_x$ and $k_y$, respectively. Taking this situation into account, we construct the ansatz
\begin{equation}
\label{eq:symansatz}
\mathbf{m}_{iQxy}(\xi)=\left(\begin{matrix} X_i \cos \left( k_{x,i} \xi\pm k_{x,i}(L+d_\text{N}/2)\right)\\ Y_i\cos \left( k_{y,i} \xi\pm k_{y,i}(L+d_\text{N}/2)\right)\end{matrix} \right) ,
\end{equation}
which, when inserted into the boundary conditions of Eqs.~\eqref{boundary condition a} and \eqref{boundary condition b}, yields
\begin{subequations}
\begin{eqnarray}
i \chi_\nu \omega_\nu + \frac{L K_S}{A} \cos \left(2\theta\right)&=k_xd\tan\left(k_xd\right),\\
i \chi_\nu \omega_\nu + \frac{L K_S}{A} \cos^2 \left(\theta\right)&=k_yd\tan\left(k_yd\right),
\end{eqnarray}
\end{subequations}
where $\nu$ continues to denote an acoustic (A) or optical (O) mode, $\nu=\text{A},\text{O}$. Depending on the sign of $K_S$ and the angle $\theta$, the resulting solutions $k_x$ and $k_y$ can become complex numbers, which implies that the modes are evanescent. Let us consider the case of $K_S>0$ and an in-plane magnetization ($\theta=\pi/2$). Although $k_y$ is unchanged by the EASA, with $LK_S/A>1\gg \chi_\nu \omega_\nu$, $k_x$ is almost purely imaginary, $\kappa=i k=K_S/A-i\omega_\nu \chi_\nu$, so that
\begin{equation}\label{eq:surfacemode}
m_{iQx}(\xi)=X\cosh(\kappa \xi \pm \kappa(d+d_\text{N}/2)).
\end{equation}
The magnetization along the $x$ direction is exponentially localized at the FI$|$N surfaces. Following the same procedure as in Sec.~\ref{symFInoEASA} for the $K_S=0$ case, we insert this solution into the dispersion relation \eqref{bulk dispersion} and extract the renormalization of the effective Gilbert damping:
\begin{equation}\label{damping EASA}
\Delta \alpha_\nu^\text{EASA}=\frac{\gamma\hbar^2 g_\perp}{2LM_Se^2} \frac{\chi_\nu}{\chi}\frac{1+\frac{\omega_H}{\omega_M}\left[1+\frac{2LK_S}{A}\right]-\frac{K_S^2}{2\pi M_S^2 A}}{1+2\frac{\omega_H}{\omega_M}-\frac{K_s^2}{2\pi M_S^2 A}}.
\end{equation}
In the presence of EASA, the damping coefficient is a tensor; thus, the effective damping of Eq.~\eqref{damping EASA} is an average, as defined in Eq.~\eqref{eq:alphadef}. This Gilbert damping enhancement may become orders of magnitude larger than the $\Delta \alpha_\text{macro}$ of Eq.~\eqref{eq:macroalpha}. For thick films, $\Delta \alpha_\text{macro} \sim L^{-1}$, whereas $\Delta \alpha_\nu ^\text{EASA}$ reaches a constant value that is inversely proportional to the localization length at the FI$|$N interface.  Note that for large EASA, the equilibrium magnetization is no longer oriented along the external field, and Eq. (29) for $\Delta \alpha^\text{EASA}_\nu$ becomes invalid. 

\subsection{Asymmetric FI Films}
Let us now consider an asymmetric system in which $L_1\neq L_2$. In this configuration, we will first consider $K_S=0$, but we will also comment on the case of a finite $K_S$ at the end of the section. Because the analytical expressions for the eigenfrequencies and damping coefficients are lengthy, we focus on the most interesting case: that in which the spin-relaxation rate is slow.  

As in the case of the symmetric films, the dispersion relation of Eq.~\eqref{matrix LLG} dictates that the wave numbers in the two layers must be the same. To satisfy the boundary equations \eqref{boundary condition}, we construct the ansatz
\begin{equation}
\label{eq:asymansatz}
\mathbf{m}_{iQxy}(\xi)=\left(\begin{matrix} X_i \cos \left( k \xi\pm k(L+d_\text{N}/2)\right)\\ Y_i\cos \left( k \xi\pm k(L+d_\text{N}/2)\right)\end{matrix} \right).
\end{equation}
The difference between this ansatz and the one for the symmetric case represented by Eq.~(\ref{eq:symansatz}) is that the magnitudes of the amplitudes, $X_i$ and $Y_i$, of the two layers, $i=1,2$, that appear in Eq.~(\ref{eq:asymansatz}) is no longer expected to be equal. 

When the two ferromagnets FI($L_1$) and FI($L_2$) are completely disconnected, the transverse wave vectors must be equivalent to standing waves, $q_{n,1}=\pi n /L_1$ and $q_{m,2}=\pi m/L_2$ in the two films, respectively, where $n$ and $m$ may be any integral numbers. Because spin pumping is weak, the eigenfrequencies of the coupled system are close to the eigenfrequencies of the isolated FIs. This finding implies that the wave vector $k$ of the coupled system is close to either $q_{n,1}$ or $q_{m,2}$. The solutions of the linearized equations of motion are then
\begin{subequations}
\begin{eqnarray}
k&=&k_{n,1}=q_{n,1}+\delta k_{n,1} \hspace{5mm} \text{or}\label{eq:wavenr1}\\
k&=&k_{m,2}=q_{m,2}+\delta k_{m,2},
\end{eqnarray}
\end{subequations}
where $\delta k_{n,1}$ and $\delta k_{m,2}$ are small corrections attributable to spin pumping and spin transfer, respectively. Here, the indices $1$ and $2$ represent the different modes rather than the layers. However, one should still expect that mode 1(2) is predominantly localized in film 1(2). In this manner, we map the solutions of the wave vectors in the coupled system to the solutions of the wave vectors in the isolated FIs. Next, we will present solutions that correspond to the $q_{n,1}$ of Eq.~\eqref{eq:wavenr1}. The other family of solutions, corresponding to $q_{m,2}$, is determined by interchanging $L_1 \leftrightarrow L_2$ and making the replacement $n \rightarrow m$. 

Inserting Eq.~\eqref{eq:wavenr1} into the boundary conditions of Eq.~\eqref{boundary condition} and linearizing the resulting expression  in the weak spin-pumping-induced coupling, we find, for the macrospin modes, 
\begin{equation}
i\omega\tilde{\chi}_{1,\text{macro}}^\text{A,O}=(L_1 \delta k_{0,1})^2,
\end{equation}
where
\begin{subequations}\label{eq:chirenorm}
\begin{eqnarray}
\tilde{\chi}_{1,\text{macro}}^\text{A}&\approx& \frac{1}{2}\frac{d_\text{N}}{l_\text{sf}}\frac{\sigma}{g_\perp l_\text{sf}}\frac{L_1}{L_1+L_2}\chi_1,\\
\tilde{\chi}_{1,\text{macro}}^\text{O}&\approx &\frac{1}{2}\frac{L_1+L_2}{L_2}\chi_1.\label{eq:chirenormb}
\end{eqnarray}
\end{subequations}
Here, $\chi_1=L_1 \hbar^2 g_\perp /4 A e^2$. Inserting this parameter into the dispersion relation of Eq.~\eqref{bulk dispersion}, we obtain the following damping renormalizations:
\begin{subequations} 
\begin{eqnarray}
\Delta \alpha_{ \text{macro}}^\text{A}&=&  \frac{\gamma\hbar^2 g_\perp}{2M_Se^2}  \frac{1}{2}\frac{d_\text{N}}{l_\text{sf}}\frac{\sigma}{g_\perp l_\text{sf}}\frac{1}{L_1+L_2},\\
\Delta \alpha_{ \text{macro}}^\text{O}&=&  \frac{\gamma\hbar^2 g_\perp}{2M_Se^2}  \frac{1}{2}\left(\frac{1}{L_1}+\frac{1}{L_2}\right) .
\end{eqnarray}
\end{subequations}
These two solutions correspond to an acoustic mode and an optical mode, respectively. The corresponding eigenvectors are $\mathbf{m}_1=\mathbf{m}_2$ for the acoustic mode and $L_1 \mathbf{m}_1=-L_2 \mathbf{m}_2$ for the optical mode. As in the symmetric case, the damping enhancement of the acoustic mode vanishes in the thin-NM limit. In this limit, the behavior of the acoustic mode resembles that of a single FI of thickness $L_1+L_2$. It is the total thickness that determines the leading-order contribution of the damping renormalization. The optical mode, however, experiences substantial damping enhancement. For this mode, the damping renormalization is the average of two separate FIs that are in contact with a perfect spin sink. The cause of this result is as follows. When there is no spin-memory loss in the NM, half of the spins that are pumped out from one side return and rectify half of the angular-momentum loss attributable to spin pumping. Because the magnetization precessions of the two films are completely out of phase, the other half of the spin current causes a dissipative torque on the opposite layer. In effect, spin pumping leads to a loss of angular momentum, and the net sum of the spin pumping across the NM and the back flow is zero. The total dissipation is not affected by spin transfer, and thus, the result resembles a system in which the NM is a perfect spin sink.

For the higher excited transverse modes, there are two scenarios, which we treat separately. I. The allowed wave number for one layer matches a wave number for the other layer. Then, for some integer $n>0$, $q_{n,1}= q_{m,2}$ for some integer $m$. In this case, we expect a coupling of the two layers. II. The allowed wave number for one layer does not match any of the wave numbers for the other layer, and thus, for some integer $n>0$, we have $q_{n,1}\neq q_{m,2}$ for all integers $m$. We then expect that the two layers will not couple. 

I. In this case, we find two solutions that correspond to acoustic and optical modes. These modes behave very much like the macrospin modes; however, as in the symmetric case, the damping renormalization is greater by a factor of 2:
\begin{equation}
\Delta \alpha^{\text{A,O}}_{n\neq 0} =2\Delta \alpha^\text{A,O}_{\text{macro}},\hspace{5mm}\text{Case I}.
\end{equation}
The eigenvectors of these coupled modes have the same form as for the macrospin modes, such that $\mathbf{m}_1=\mathbf{m}_2$ and $L_1 \mathbf{m_1}=-L_2 \mathbf{m_2}$ for the acoustic and optical modes, respectively.

II. In this case, the two layers are completely decoupled. To the leading order in $d_\text{N}/l_\text{sf}$, we find
\begin{equation}
\Delta \alpha_{n\neq 0} =\frac{\gamma\hbar^2 g_\perp}{2L_1M_Se^2}  ,\hspace{5mm}\text{Case II},
\end{equation}
for all modes that correspond to excitations in FI1. The damping renormalization is thus half that of the FI($L_1$)$|$N($l_\text{sf}=0$) system.\cite{Kapelrud:prl13} This result can be explained by the zero loss of spin memory in the NM. Although half of the spins are lost to the static FI2, half of the spins return and rectify half of the dissipation attributable to spin pumping. The amplitudes of these modes are strongly suppressed in FI2 (or FI1, upon the interchange of FI1 $\leftrightarrow$ FI2), such that $\mid \mathbf{m}_2\mid /\mid \mathbf{m}_1 \mid \sim \omega \chi_2$. 

Finally, let us discuss the case in which EASA is present. In the limit $K_S L_i/A \gg 1$, the excitation energies of the surface modes are independent of the FI thicknesses. However, the surface modes do not behave like the macrospin modes for the asymmetric stack. The excitation volume of these modes is determined by the decay length $A/K_S$ in accordance with Eq.~\eqref{eq:surfacemode}. This finding is in contrast to the result for the macrospin modes, where the excitation volume spans the entire FI. Thus, the surface modes couple in the same manner as in the symmetric case. With a good experimental control of surface anisotropy, the coupling of the surface modes is thus robust to thickness variations. The higher excited transverse modes, in the presence of EASA, have thickness-dependent frequencies, which means that these modes behave similarly to the $n >0$ modes in the $K_S=0$ case. 

\section{Numerical results\label{numerical} }
When the spin-wave wavelength becomes comparable to the film thickness, the dipolar field becomes a complicated function of the wavelength. We study the properties of the system in this regime by numerically solving the linearized equations of motion \eqref{matrix LLG} with the boundary conditions \eqref{boundary condition}. We use the method presented in Ref.~\onlinecite{Kapelrud:prl13}, which solves the spin-wave excitation spectrum for an FI$|$N system, and extend this approach to the present trilayer system. The physical parameters used in the numerical calculations are listed in Table~\ref{physicalparameters}. We investigate two geometries: I. the BWMSW geometry, in which the spin wave propagates parallel to the external field, and II. the MSSW geometry, in which the spin wave propagates perpendicular to the external field. 

To calculate the renormalization of the Gilbert damping, we perform one computation without spin pumping and one computation with spin pumping, in which the intrinsic Gilbert damping is excluded. Numerically, the renormalization can then be determined by calculating $\Delta \alpha=\alpha \text{Im}[\omega^{(\text{SP})}]_{\alpha=0}/\text{Im}[\omega^{(0)}]$, where $\omega^{(0)}$ is the eigenfrequency obtained for the computation without spin pumping and $\omega^{(\text{SP})}$ is the frequency obtained for the computation with spin pumping.\cite{Kapelrud:prl13}

\begin{table}[h!]

\centering

\caption{Physical parameters used in the numerical calculations}

\begin{tabular}{@{} lcc @{}}

\hline

Constant & Value & Units   \\

\hline

$g_\perp$ & $^a3.4 \cdot 10^{15}$ &$\mathrm{cm}^{-2}e^2/h$ \\

$\sigma$ & $^b5.4\cdot 10^{17}$ &$\mathrm{s}^{-1}$ \\

$4\pi M_S$ & $^c1750$& $\mathrm{G}$  \\

$A$ & $^c3.7 \cdot 10^{-7}$  & $\mathrm{erg/cm}$ \\

$H_\text{int}$ & $0.58\cdot 4\pi M_S$ & \\

$\alpha$ & $^c3 \cdot 10^{-4}$ &  \\
$K_S$ & $ 0, \hspace{1mm}^d0.05$& $\mathrm{erg/cm^2}$ \\

\hline
\hline
\label{physicalparameters}
\end{tabular}

\vspace{1mm}

a) Ref.~[\onlinecite{Junfleisch:apl13}], b) Ref.~[\onlinecite{Giancoli:book84}], c) Ref.~[\onlinecite{Serga:jphysd}]\\
d) Reported to be in the range of $0.1-0.01~\mathrm{erg/cm^2}$ in Ref.~[\onlinecite{Xiao:prl12}]

\label{tab:constants}
\end{table}

\subsection{BVMSW}
\begin{figure}[!ht] 
    \includegraphics[width=0.42\textwidth]{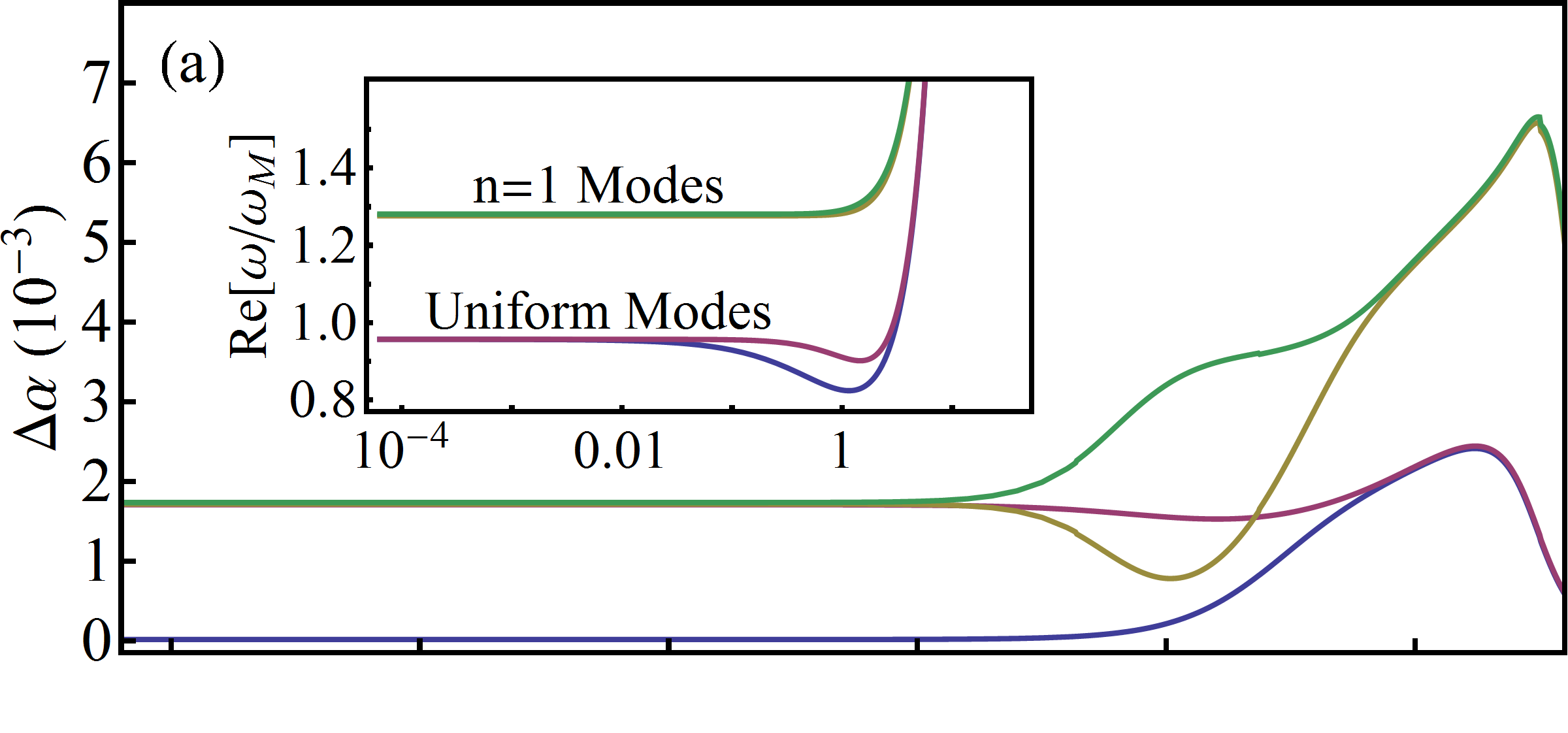}\\ \vspace{-2mm}
  \includegraphics[width=0.42\textwidth]{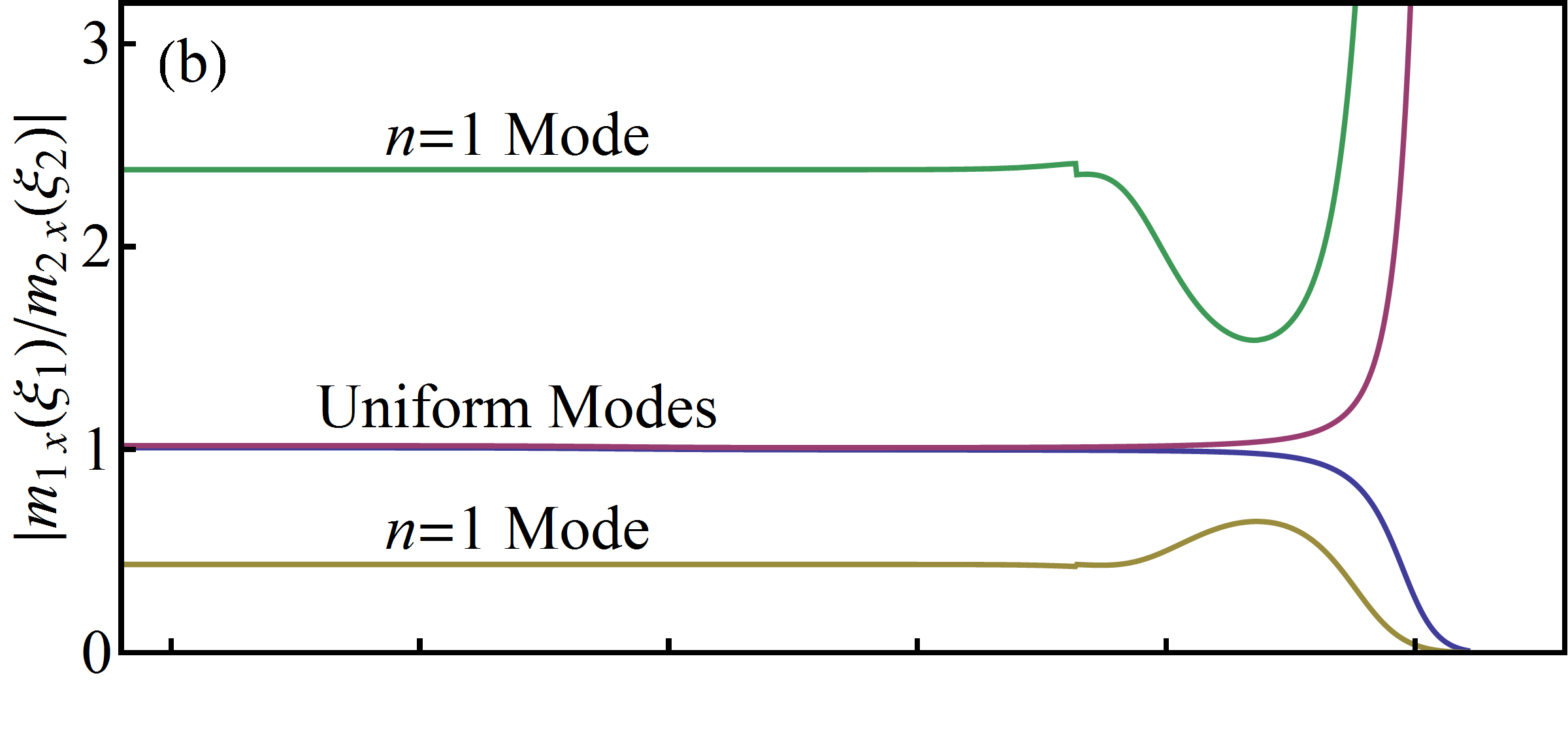}\\ \vspace{-2mm}
     \includegraphics[width=0.42\textwidth, trim=12mm 0mm 0mm 0]{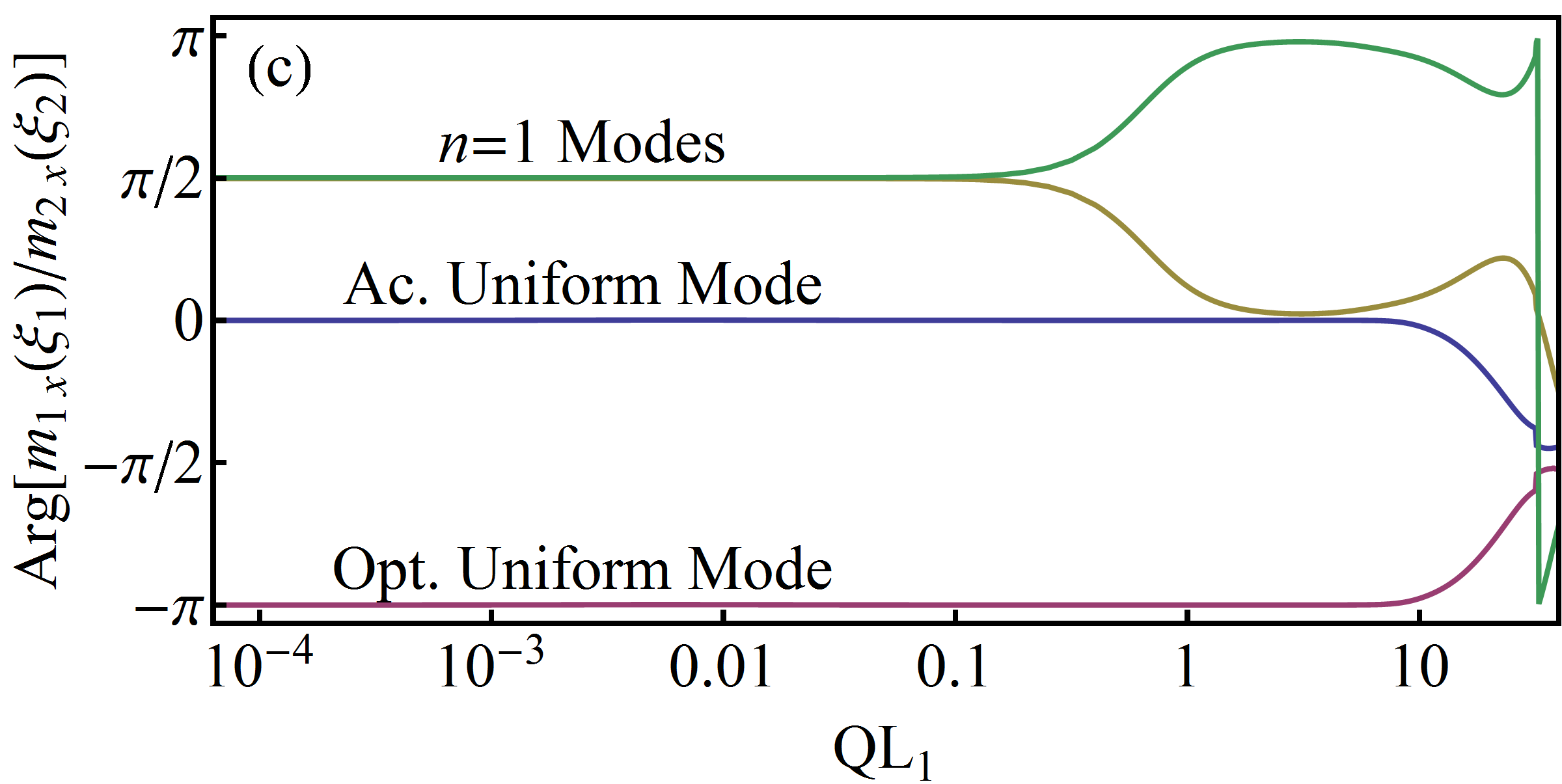}\hspace{30mm}
	\caption{(Color online) FI(100nm)$|$N(50nm)$|$FI(101nm): a) Spin-pumping-enhanced Gilbert damping $\Delta \alpha$ as a function of $QL_1$ of the uniform modes and the $n=1$ modes. The inset presents the corresponding dispersion relation. b) Relative phase and c) amplitude between the out-of-plane magnetizations along $x$ at the edges of FI1$|$N and FI2$|$N. The apparent discontinuity in the green line in c) appears because the phase is defined on the interval $-\pi$ to $\pi$. \label{fig:100_50_100}}
\end{figure}
Let us first discuss the BVMSW geometry. The coupling of the uniform modes in the two films is robust; it is not sensitive to possible thickness asymmetries. In contrast, at $Q=0$, the sensitivity to the ratio between the thickness and the rather weak dynamic coupling attributable to spin pumping implies that the coupling of the higher transverse modes in the two bilayers is fragile.

Small asymmetries in the thicknesses destroy the coupling. This effect can best be observed through the renormalization of the damping. However, we will demonstrate that a finite wave number $Q$ can compensate for this effect such that the higher transverse modes also become coupled. To explicitly demonstrate this result, we numerically compute the real and imaginary parts of the eigenfrequencies of a slightly asymmetric system,  FI(100nm)$|$N(50nm)$|$FI(101nm) with $l_\text{sf}=350$ nm. The asymmetry between the thicknesses of the ferromagnetic insulators is only 1\%. The surface anisotropy is considered to be small compared with the ratio $L_i/A$, and we set $K_S=0$. 

In Fig.~\ref{fig:100_50_100}, the numerical results for the effective Gilbert damping, the dispersion of the modes, and the relative phase and amplitude between the magnetizations in the two FIs are presented. As observed in the relative phase results depicted in Fig.~\ref{fig:100_50_100}(c), the two uniform modes in widely separated FIs split into an acoustic mode and an optical mode when the bilayers are coupled via spin pumping and spin transfer. Figure~\ref{fig:100_50_100}(a) also demonstrates that the acoustic mode has a very low renormalization of the Gilbert damping compared with the optical mode. Furthermore, there is no phase difference between the two modes with a transverse node ($n=1$) in  Fig. \ref{fig:100_50_100}(a), which indicates that the modes are decoupled. These $n=1$ modes are strongly localized in one of the two films; see Fig.~\ref{fig:100_50_100}(b). For small $QL_1$, Fig.~\ref{fig:100_50_100}(a) demonstrates that these modes have approximately the same renormalization as the optical mode, which is in agreement with the analytical results. Because the magnetization in the layer with the smallest amplitude is only a response to the spin current from the other layer, the phase difference is $\pi/2$ (Fig.~\ref{fig:100_50_100}(b)). When $Q$ increases, the dipolar and exchange interactions become more significant. The interlayer coupling is then no longer attributable only to spin pumping but is also caused by the long-range dipole-dipole interaction. This additional contribution to the coupling is sufficient to synchronize the $n=1$ modes. The relative amplitude between the two layers then becomes closer to $1$ (see Fig.~\ref{fig:100_50_100}(b)). Again, we obtain an acoustic mode and an optical $n=1$ mode, which can be observed from the phase difference between the two layers in Fig.~\ref{fig:100_50_100}(c). The spin-pumping-induced coupling only occurs as long as the effective spin-diffusion length $\tilde{l}_\text{sf}$ is large or on the order of $d_\text{N}$. Once this is no longer the case, the modes rapidly decouple, and the system reduces to two separate FI$|$N systems with a relatively weak interlayer dipole coupling. In the limit of large $QL_1$, the exchange interaction becomes dominant. The energy of the wave is then predominantly attributable to the momentum in the longitudinal direction, and the dynamic part of the magnetization goes to zero at the FI$|$N interfaces, causing the renormalization attributable to spin pumping to vanish.\cite{Kapelrud:prl13}  

We also note that the dispersion relation depicted in the inset of Fig.~\ref{fig:100_50_100}(a) reveals that the acoustic mode (blue line) exhibits a dip in energy at lower $QL_1$  than does the optical mode (red line). We suggest that this feature can be understood as follows: The shift in the position of the energy dip can be interpreted as an increase in the effective FI thickness for the acoustic mode with respect to that for the optical mode. When $\tilde{l}_\text{sf}$ is larger than the NM thickness, the uniform mode behaves as if the NM were absent and the two films were joined. This result indicates that the dispersion relation for the acoustic mode exhibits frequency behavior as a function of $Q\tilde{L}/2$, where the effective total thickness of the film is $\tilde{L}=L_1+L_2$. The optical mode, however, ``sees'' the NM and thus behaves as if $\tilde{L}=L_1$. Consequently, the dip in the dispersion occurs at lower $QL_1$  for the acoustic mode than for the optical mode. 

\subsection{MSSW}
Finally, let us study the dynamic coupling of magnetostatic surface spin waves (MSSWs). We now consider a perfectly symmetric system, FI(1000 nm)$|$N(200 nm)$|$FI(1000 nm), with $l_\text{sf}=350~\mathrm{nm}$. For such thick films, surface anisotropies may play an important role. We therefore discuss a case in which we include a surface anisotropy of $K_S=0.05~\mathrm{erg/cm^2}$. According to the analytical result presented in Eq. (28), the lowest-energy modes with $QL_1\ll 1$ are exponentially localized at the FI$|$N surfaces, with a decay length of $A/K_S\sim 200~\mathrm{nm}$. 

\begin{figure}[!ht]  
    \includegraphics[width=0.43\textwidth,trim=4mm 0 -2mm 0]{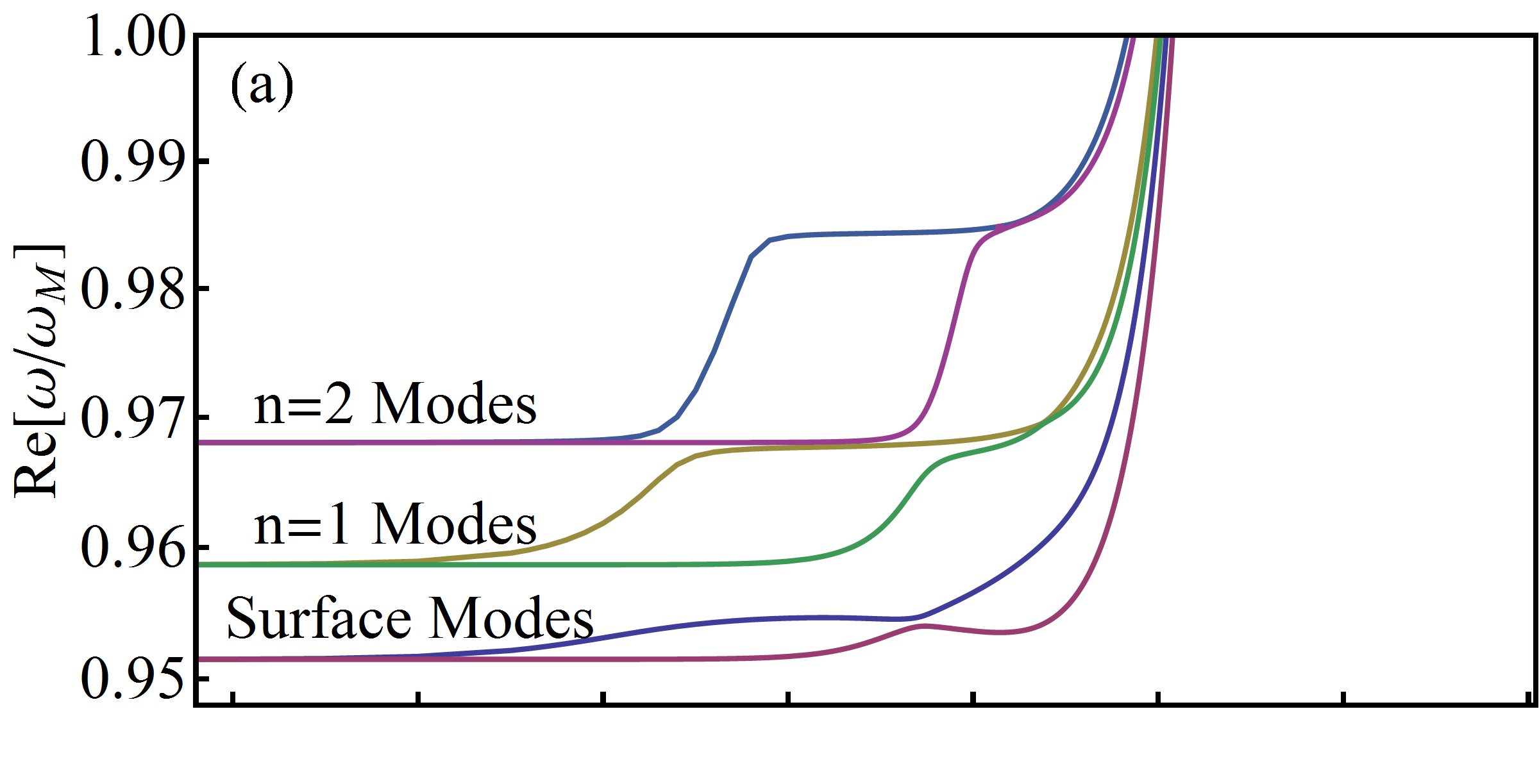}\hspace{10mm}\\ \vspace{-2.9mm}
      \includegraphics[width=0.45\textwidth]{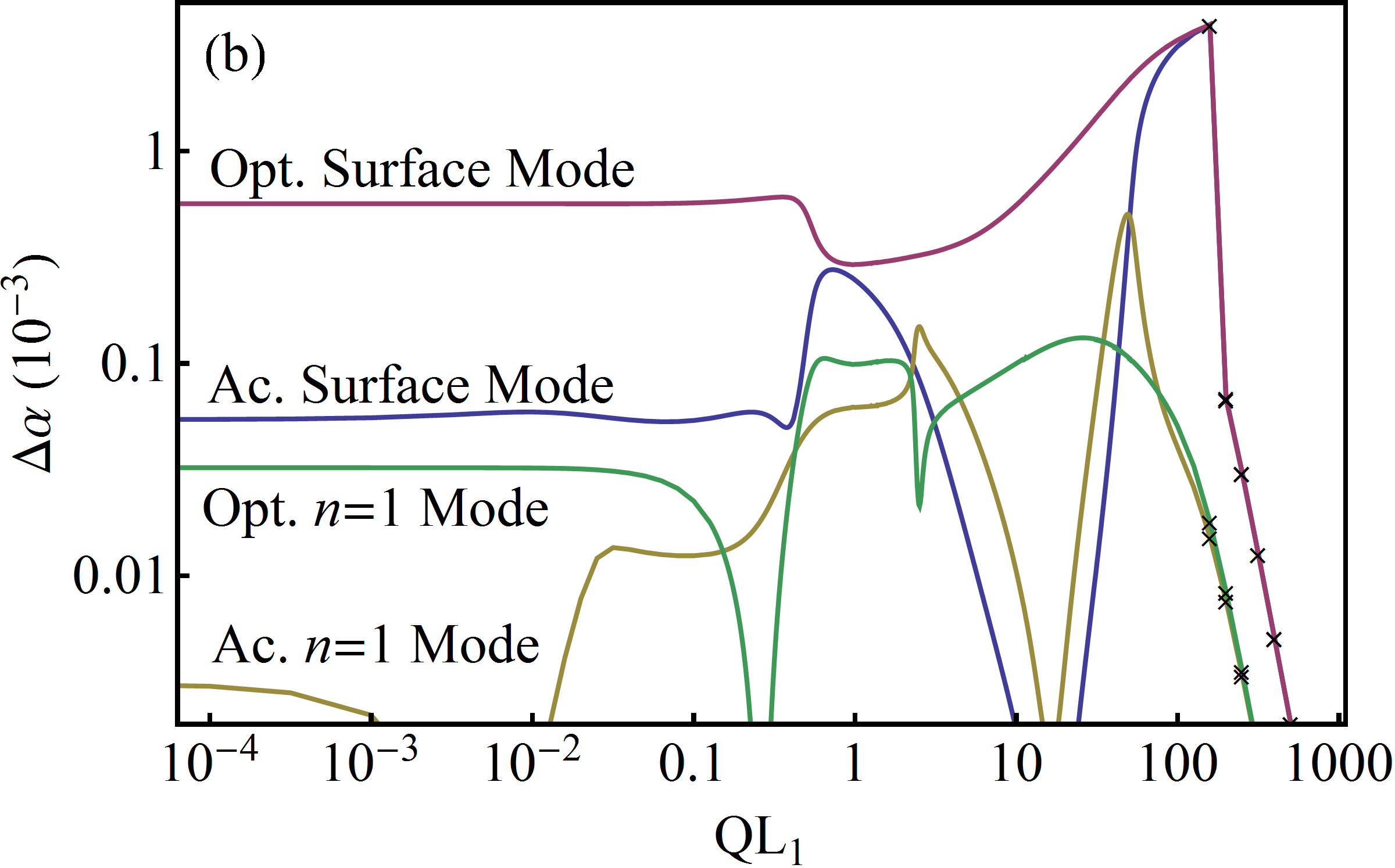}
    \caption{(Color online) {FI(1000nm)$|$N(200nm)$|$FI(1000nm)} $l_\text{sf}=350~\mathrm{nm}$, $K_S=0.05 ~\mathrm{erg/cm^2}$: a) The dispersion relation as a function of $QL_1$ for the six lowest positive-real-part modes. b) The renormalization of the damping attributable to spin pumping for the four lowest modes with frequencies with positive real parts as a function of $QL_1$. 
At large $QL_1$, the computation becomes increasingly demanding, and the point density of the plot becomes sparse. We have therefore individually marked the plotted points in this region. \label{fig:1000_200_1000_EASA}}
\end{figure}

We now compute the eigenfrequencies, $\omega$, as a function of the wave vector in the range $10^{-4}<QL_1<10^3$. In Fig.~\ref{fig:1000_200_1000_EASA}(a), we present the real part of the frequency for the six lowest-energy modes with a positive real part, and in Fig. 4(b), we present the corresponding renormalizations of the Gilbert damping for the four lowest-energy modes. The dispersion relations indicate  that the mode pairs that are degenerate at $QL_1\ll1$ rapidly split in energy when $QL_1$ approaches $10^{-2}$. Strong anticrossings can be observed between the $n=1$ and $n=2$ modes. Such anticrossings are also present between the surface mode and the $n=1$ mode; they are almost too strong to be recognized as anticrossings. The enhanced damping renormalizations exhibit very different behavior for the different modes. We recognize the large-$\Delta \alpha$ mode of one pair as the surface optical mode and the low-$\Delta \alpha$ mode as the volume $n=1$ acoustic mode. Without EASA, the anticrossings in Fig.~\ref{fig:1000_200_1000_EASA}(a) would become crossings. The lowest-energy modes at $QL_1\ll 1$ would then cut straight through the other modes. In the case considered here, this behavior is now observed only as steep lines at $QL_1\sim 0.05$ and at $QL_1\sim 0.5$.  

When $Q$ is increased, the effective spin-diffusion length decreases (see Eq.~\eqref{eq:lsfeff}), which reduces the spin-pumping-induced coupling between the modes at large $Q$. When $QL_1\sim 100$, the coupling becomes so weak that the two FIs decouple. This phenomenon can be observed from the behavior of $\Delta \alpha$ in Fig.~\ref{fig:1000_200_1000_EASA}(b), where the damping of the acoustic modes become the same as for the optical modes.

\begin{figure}[!ht]  
    \includegraphics[width=0.41\textwidth,trim=10mm 0 0mm 0]{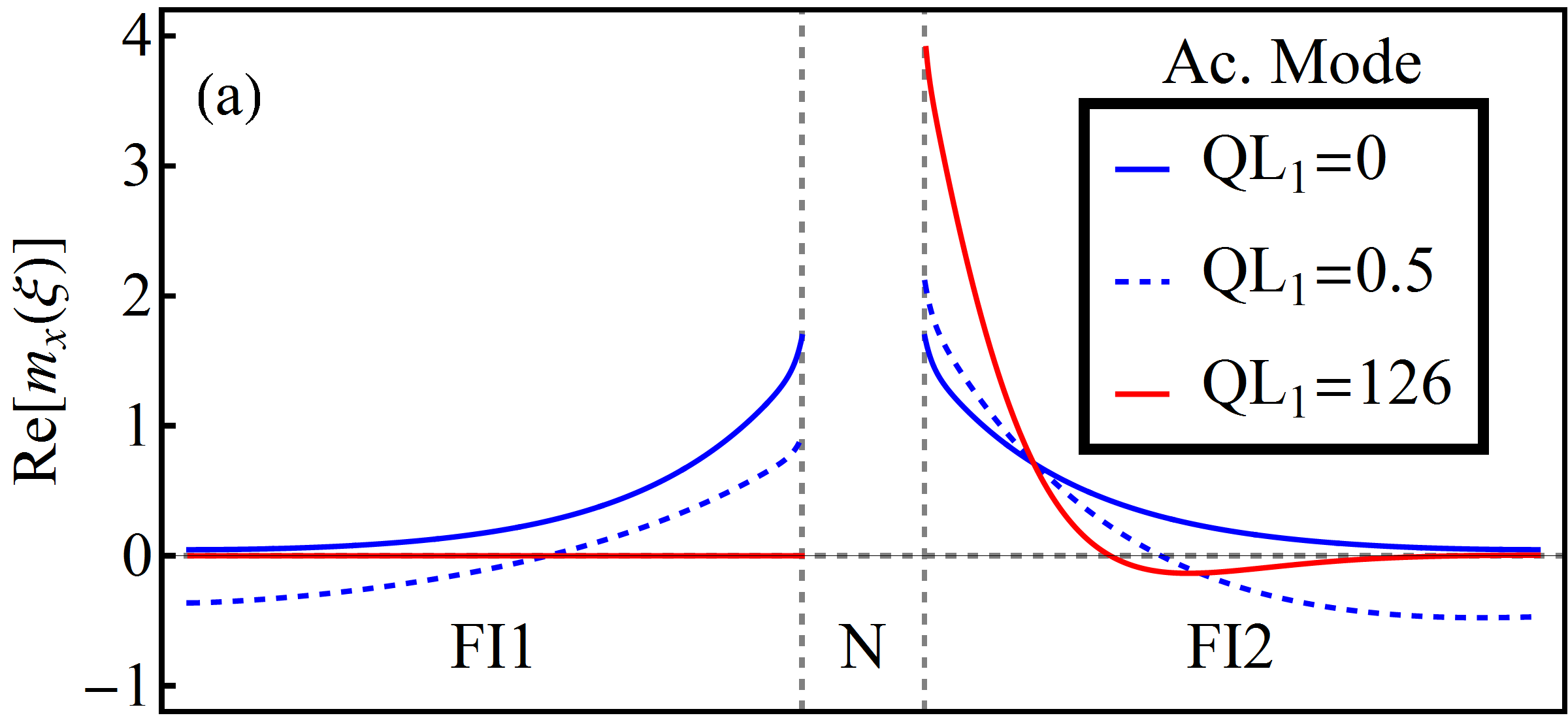}\hspace{10mm}\\
    \includegraphics[width=0.45\textwidth]{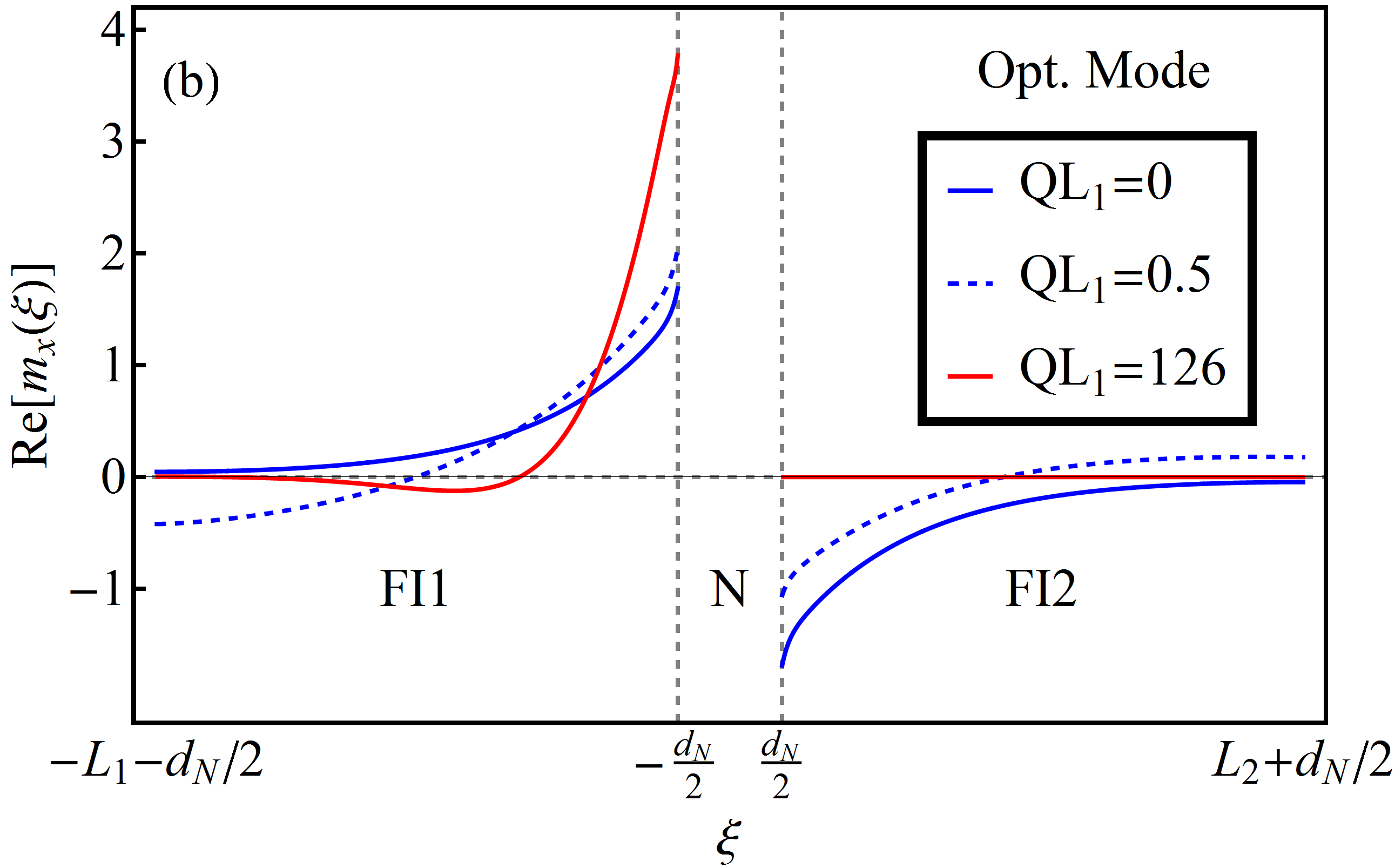}
    \caption{(Color online){FI(1000nm)$|$N(200nm)$|$FI(1000nm)}, $l_\text{sf}=350 \mathrm{nm}$, $K_S=0.05~\mathrm{erg/cm^2}$: a) and b) present the real parts of the $x$ components of the out-of-equilibrium magnetization vectors for the acoustic and optical surface modes, respectively, for several values of $QL_1$. For values of $QL_1\gtrsim 1$, the modes decouple and become localized in one of the two layers. For large values of $QL_1\sim 100$, the two modes are strongly localized at one of the two FI$|$N interfaces, which correspond to the peaks in the damping that are apparent in Fig. \ref{fig:1000_200_1000_EASA}(b). \label{fig:eigenvectors}}
\end{figure}

In the MSSW geometry, an isolated FI has  magnetostatic waves that are localized near one of the two surfaces, depending on the direction of propagation with respect to the internal field.\cite{Serga:jphysd} Asymmetries in the excitation volume are therefore also expected for the trilayer in this geometry. In Fig.~\ref{fig:eigenvectors}, we present the eigenvectors of the surface modes as functions of the transverse coordinate $\xi$ for increasing values of the wave vector $Q$. At $QL_1=0.5$, the modes have already begun to exhibit some asymmetry. Note that the renormalization of the damping observed in Fig.~\ref{fig:1000_200_1000_EASA}(b) is approximately one order of magnitude larger than the intrinsic Gilbert damping for the optical mode and that the damping of any one mode may vary by several orders of magnitude as a function of $QL_1$.\cite{Kapelrud:prl13} Therefore, these effects should be experimentally observable. The greatest damping occurs when the two layers are completely decoupled; see Figs. Fig.~\ref{fig:1000_200_1000_EASA}(b) and \ref{fig:eigenvectors}. Because the damping of the optical mode is equivalent to that of a system with a perfect spin sink, one might expect that the greatest damping should occur for this mode. However, the large localization, which is achieved only at large $QL_1$, in combination with the vanishing of the effective spin-diffusion length leads to damping that is much greater than that of the synchronized optical mode. 

\section{Conclusions}
\label{conclusion}
We investigated the dynamic coupling of spin-wave excitations, which are present in single FI thin films, primarily through spin pumping and spin transfer but also through the dynamic demagnetization field created when two FI thin films are in contact via an NM layer. Because of this coupling, the modes are split into acoustical and optical excitations. When the NM is thin compared with $l_\text{sf}$, the renormalization of the Gilbert damping vanishes for the acoustic modes, whereas for the optical modes, the renormalization is equally as large as for a single-FI$|$N system in which the NM is a perfect spin sink. A spin current pumped by a travelling magnetic wave has a wavelength of equal magnitude, which leads to traversal paths across the NM that are longer than the thickness of the NM. Consequently, the spin-memory loss is greater for short-wavelength spin currents. This phenomenon leads to an effective spin-diffusion length in the NM that decreases for increasing values of $Q$. As a result, the dynamic coupling strength is reduced for short-wavelength spin waves. At some critical value of $Q$, the coupling becomes so weak that the acoustic-~and optical-mode configurations are lost in favor of modes that are localized in one of the two FIs. At these values of $Q$, the interlayer dipole coupling is also dominated by the intralayer exchange coupling. For these high-wave-number modes, the system behaves similar to two separate FI$|$N$(l_\text{sf}=0)$ systems. 

When the two films are of different thicknesses, the exchange energies of the higher-order transverse $n>1$ modes differ between the two layers. Because of the relatively small coupling attributable to spin pumping, the synchronization of these modes at $QL_1\ll1$ requires that the FI thicknesses be very similar. A small asymmetry breaks the synchronization; however, for larger $QL_1\sim 1$, the modes can again become coupled through interlayer dipole interaction. This coupling arises in addition to the spin-pumping- induced coupling. For even larger $Q$, the effective spin-diffusion length becomes small, and the coupling attributable to spin pumping vanishes. The relatively small dipole coupling alone is not sufficient to couple the modes when there is a finite difference in film thickness , and the synchronization breaks down. 

Depending on the quality of the interface between the FIs and the strength of the spin-orbit coupling in the NM , additional effective surface fields may be present because of surface anisotropy energies. For the EASA case, the lowest-energy modes are localized at the FI$|$N surfaces. These modes couple in the same manner as the macrospin modes. For films that are much thicker than the decay length $A/K_S$, the energies of the surface modes do not depend on the film thickness. Consequently, the coupling of these modes is independent of the thickness of the two FIs. Similar to the simpler FI$|$N system, the damping enhancement may attain values as high as an order of magnitude larger than the intrinsic Gilbert damping. However, in the trilayer system, the presence of both acoustic and optical modes results in large variations in the effective damping within the same physical sample. Because of this wide range of effective damping, which spans a difference in $\Delta \alpha$ of several orders of magnitude as a function of $Q$, we suggest that trilayer modes should be measurable in an experimental setting.

With more complicated FI structures in mind, we believe that this work may serve as a guide for experimentalists. The large variations in effective damping for different modes make the magnetic properties of the system  detectable both with and without EASA. For spin waves, dipole-dipole interactions assist spin pumping in interlayer synchronization, which may facilitate the design of future spintronic devices.

\acknowledgements
We acknowledge support from InSpin 612759 and the Research Council of Norway, project number 216700.


\appendix
\section{Dipole Tensor}\label{app: dipole tensor}
The dipole tensor in the $\zeta \eta \xi$ coordinate system, $\hat{G}(\xi)$ from Eq.~\eqref{kalinikos} can be rotated by the $xyz$ coordinate system with the rotation matrix
\begin{equation}
R=\left(\begin{matrix} s_\theta & -c_\theta s_\theta & -c_\theta c_\phi \\
0 & c_\phi & -s_\phi \\
c_\theta & s_\theta s_\phi & s_\theta c_\phi \end{matrix} \right),
\end{equation}
where we have introduced the shorthand notation $s_\theta \equiv \sin \theta$, $c_\theta \equiv \cos \theta$ and so on. We then get that
\begin{widetext}(*
\begin{equation}
\begin{split}
\hat{\mathcal{G}}_{xyz}=&R \hat{G} R^T\\
=&\left( \begin{matrix}
s_\theta^2 G_{\xi \xi} - c_\phi s_{2\theta} G_{\xi\zeta}+c_\theta^2 c_\phi^2 G_{\zeta \zeta} & -s_\phi s_\theta G_{\xi \zeta} + s_\phi c_\phi c_\theta G_{\zeta \zeta} & s_\theta c_\theta G_{\xi \xi} - s_\theta c_\theta c_\phi^2 G_{\zeta \zeta} + c_\phi (s_\theta^2-c_\theta^2)G_{\xi \zeta}\\
-s_\phi s_\theta G_{\xi \zeta} + s_\phi c_\phi c_\theta G_{\zeta \zeta} & s_\phi^2 G_{\zeta \zeta} & -s_\phi c_\theta G_{\xi \zeta} + s_\phi s_\theta c_\phi G_{\zeta \zeta} \\
s_\theta c_\theta G_{\xi \xi} -s_\theta c_\theta c_\phi^2 G_{\zeta \zeta} + c_\phi (s_\theta^2-c_\theta^2)G_{\xi \zeta} & - s_\phi c_\theta G_{\xi \zeta} +s_\phi s_\theta c_\phi G_{\zeta \zeta} & c_\theta^2 G_{\xi \xi} + s_{2\theta} c_\phi G_{\xi \zeta} + c_\phi^2 s_\theta^2 G_{\zeta \zeta} 
\end{matrix}\right).
\end{split}
\end{equation}
Because we work in the linear respons regime the equilibrium magnetization should be orthogonal to the dynamic deviation, $\mathbf{m}_i\cdot \hat{\mathbf{z}}=0$,  it is therefor sufficient to only keep the $xy$ part of $\hat{\mathcal{G}}_{xyz}$. We then find 
\begin{equation}
\begin{split}
\hat{\mathcal{G}}_{xy}=\left(\begin{matrix}
s_\theta^2 G_{\xi \xi} - c_\phi s_{2\theta} G_{\xi\zeta}+c_\theta^2 c_\phi^2 G_{\zeta \zeta} & -s_\phi s_\theta G_{\xi \zeta} + s_\phi c_\phi c_\theta G_{\zeta \zeta}  \\
-s_\phi s_\theta G_{\xi \zeta} + s_\phi c_\phi c_\theta G_{\zeta \zeta} & s_\phi^2 G_{\zeta \zeta} \end{matrix}\right).
\end{split}
\end{equation}
\end{widetext}

\section{Spin Accumulation}\label{app: spin acc}
The functions $\Gamma_1(\xi)$ and $\Gamma_2(\xi)$ are taken directly from Ref. \cite{Skarsvaag:14}, and modified to cover the more complicated magnetic texture model. We then have
\begin{equation}
\begin{split}
\Gamma_1\left(\xi\right)\equiv \frac{\cosh \left(\xi/\tilde{l}_\text{sf}\right)}{\cosh \left(\xi/\tilde{l}_\text{sf}\right)+\sigma \sinh \left(\xi/\tilde{l}_\text{sf}\right)/2g_\perp \tilde{l}_\text{sf}},\\
\Gamma_2\left(\xi\right)\equiv \frac{\sinh \left(\xi/\tilde{l}_\text{sf}\right)}{\sinh \left(\xi/\tilde{l}_\text{sf}\right)+\sigma \cosh \left(\xi/\tilde{l}_\text{sf}\right)/2g_\perp \tilde{l}_\text{sf}}.
\end{split}
\end{equation}
For $Ql_\text{sf}\gg 1$ the effective spin diffusion length becomes short, $\Gamma_1\rightarrow 1$ and $\Gamma_2\rightarrow 0$ at the FI$|$N interfaces.

\section{Effective spin diffusion length}\label{app: lsf}
The diffusion in the NM reads
\begin{equation}\label{eq:diffusioneq}
\partial_t \boldsymbol{\mu}_S=D\nabla^2\boldsymbol{\mu}_S-\frac{1}{\tau_\text{sf}} \boldsymbol{\mu}_S,
\end{equation}
where $D$ is the diffusion constant and $\tau_\text{sf}$ is the spin flip relaxation time. We assume that the FMR frequency is much smaller than the electron traversal time, $D/d_\text{N}^2$, and the spin-flip relaxation rate, $1/\tau_\text{sf}$.\cite{Skarsvaag:14} This means the LHS of Eq.~\eqref{eq:diffusioneq} can be disregarded. In linear response the spin accumulation, which is a direct consequence of spin pumping, must be proportional to the rate of change of magnetization at the FI$|$N interfaces. We do the same Fourier transform, as for the magnetization, so that $\boldsymbol{\mu}\sim \exp \left\{i\left(\omega t-Q \zeta\right)\right\}$. The spin diffusion equation then takes the form
\begin{equation}
\partial_\xi ^2\boldsymbol{\mu}_S=\left(Q^2+\frac{1}{D \tau_\text{sf}}\right)\boldsymbol{\mu}_S.
\end{equation}
The spin diffusion length is then $l_\text{sf}=\sqrt{D \tau_\text{sf}}$, and by introducing the effective spin diffusion length $\tilde{l}_\text{sf}=l_\text{sf}/\sqrt{1+\left(Ql_\text{sf}\right)^2}$ one gets
\begin{equation}
\partial_\xi^2 \boldsymbol{\mu}_S=\frac{1}{\tilde{l}_\text{sf}^2}\boldsymbol{\mu}_S.
\end{equation}
\end{document}